\documentclass[twocolumn]{autart}    

\usepackage{amsmath}
\usepackage{algorithm,algpseudocode}
\usepackage{amssymb}
\usepackage{color}
\usepackage{enumerate}
\usepackage{wrapfig}
\usepackage{graphicx}          
\usepackage[dvips]{epsfig}    
\usepackage[round, sort]{natbib}
\usepackage{hyperref}
\hypersetup{
	colorlinks = true,
	citecolor = cyan,
}
\usepackage{url}

\newcounter{MYtempeqncnt}

\newtheorem{theorem}{Theorem}[section]

\newtheorem{remark}{Remark}[section]
\newtheorem{definition}{Definition}[section]
\newtheorem{example}{Example}[section]

\newtheorem{assumption}{Assumption}[section]

\newtheorem{problem}{Problem}

\allowdisplaybreaks

\begin{document}
		\begin{frontmatter}
			
			\title{ Data-driven Control Against False Data Injection Attacks} 
			
			\thanks[footnoteinfo]{The work was supported 
				in part by the National Natural Science Foundation of China under Grants U23B2059,  62173034, 61925303, 62025301, 62088101, 
				the China Scholarship Council under Grants 20206120009, 202206030127,
				and the Fundamental Research Funds for the Central Universities. \emph{(Corresponding author: Gang Wang.)}}
			\thanks[footnoteinfo]{This paper was not presented at any IFAC 
				meeting. }
		
		\author[Bit,Tcic]{Wenjie Liu}\ead{liuwenjie@bit.edu.cn}, 
		\author[Rug]{Lidong Li}\ead{l.li@rug.nl},   
		\author[Bit,Tcic]{Jian Sun}\ead{sunjian@bit.edu.cn},            
		\author[Bit,Tcic]{Fang Deng}\ead{dengfang@bit.edu.cn}, 
		\author[Bit,Tcic]{Gang Wang}\ead{gangwang@bit.edu.cn}, 
		\author[Bit,Tongji]{Jie Chen}\ead{chenjie@bit.edu.cn} 
		
		\address[Bit]{State Key Lab of Autonomous Intelligent Unmanned Systems, School of Automation,\\ Beijing Institute of Technology, Beijing 100081, China}  
		\address[Rug]{Engineering and Technology Institute, University of Groningen, 9747AG, The Netherlands}             
		\address[Tcic]{Beijing Institute of Technology Chongqing Innovation Center, Chonqing 401120, China}     
		\address[Tongji]{Department of Control Science and Engineering, Tongji University, Shanghai 201804, China}        

		\maketitle

		\begin{abstract}
			\indent	The rise of cyber-security concerns has brought significant attention to the analysis and design of cyber-physical systems (CPSs). Among the various types of cyberattacks, denial-of-service (DoS) attacks and false data injection (FDI) attacks can be easily launched and have become prominent threats. While resilient control against DoS attacks has received substantial research efforts, countermeasures developed against FDI attacks have been relatively limited, particularly when explicit system models are not available.	
			To address this gap, the present paper focuses on the design of data-driven controllers for unknown linear systems subject to FDI attacks on the actuators, utilizing input-state data. To this end, a general FDI attack model is presented, which imposes minimally constraints on the switching frequency of attack channels and the magnitude of attack matrices. A dynamic state feedback control law is designed based on offline and online input-state data, which adapts to the channel switching of FDI attacks. This is achieved by solving two data-based semi-definite programs (SDPs) on-the-fly to yield a tight approximation of the set of subsystems consistent with both offline clean data and online attack-corrupted data. It is shown that under mild conditions on the attack, the proposed SDPs are recursively feasible and controller achieves exponential stability.
			Numerical examples showcase its effectiveness in mitigating the impact of FDI attacks.
		\end{abstract}

	\end{frontmatter}

	\section{Introduction}\label{sec:intro}

	\emph{Background and motivation:}
	Although the widespread applications of cyber-physical systems (CPSs) in modern industrial processes offer superior efficiency, performance, and scalability \cite{lee2008cyber, pasqualetti2013attack}, their cyber and networked aspects also introduce vulnerabilities that can be exploited through attacks, resulting in devastating damages, financial losses, and even harm to human lives \cite{luo2023secure}.
	According to Jerome Powell, Chairman of the Federal Reserve of the United States, cyberattacks have emerged as the foremost peril to the global financial system \cite{kotidis2022cyberattacks}. The COVID-19 pandemic and other recent crises have further amplified this risk, resulting in an upswing of cybercriminal activities, particularly ransomware attacks like the Kaseya VSA supply chain ransomware attack in 2021 \cite{ransom2021robert}. Remarkably, cybercrime's economic toll is estimated at 1\% of the world's GDP, equivalent to a staggering $\$1,000$ billion annually. This alarming metric emphasizes the exigent requirement for resilient cybersecurity measures and novel strategies to confront cyber threats. 

	
	\emph{Literature review:}
	In general, cyberattacks can be categorized based on their target: disrupting the availability or compromising the integrity of transmitted information \cite{SminSecure}. The former, often manifested as communication interruptions caused by malicious jammers or routers, commonly employs denial-of-service (DoS) techniques \cite{hou2022deep}. 
	Resilient control strategies have been developed since 2015, in e.g. \cite{PersisInput} to ensure acceptable performance even in the presence of such attacks. Noteworthy contributions in this area can be found in  \cite{FengResilient,wakaiki2019stabilization,Liu2021data,shi2022quantized} and their associated references. 
	
	On the other hand, the attackers aim to compromise data integrity by manipulating transmitted packets using false data injection (FDI) attacks. These attacks involve eavesdropping on authentic data and injecting false information, silently causing damage to the system performance without detection \cite{cheng2019event}. 
	Considerable efforts have been dedicated to designing detectors capable of raising alerts when anomalies are detected, ranging from model-based methods to data-driven methods, in e.g., \cite{pasqualetti2013attack,mo2013resilient,wu2020optimal,krishnan2021data}. 
	However, striking a balance between enhancing sensitivity to cyberattacks and minimizing false alarms during normal operation remains a challenge \cite{bai2017data}. 
	With the increasing sophistication and intelligence of attack strategies, the risk of undetectable attacks escalates, leading to devastating performance degradation. 
	The recent focus has shifted from dealing with individual attacks to maintaining robustness in the presence of these attacks; see e.g., \cite{anand2022risk,murguia2020security,hashemi2022co,wu2019optimalswitching}.

	Nonetheless, there are notable limitations with these existing results. 
	Primarily, all the aforementioned studies require an explicit system model, which is difficult to obtain in real-world scenarios, see e.g. \cite{Coulson2019data,persis2020data,rotulo2021online,li2023data}. 
	Moreover, to pose constraints on the power of the attacker, it was assumed in \cite{anand2022risk,murguia2020security,hashemi2022co} that both the attacking gain matrix and the injected data are upper bounded. 
	Although  \cite{wu2019optimalswitching} has relaxed this assumption by only constraining the attacking gain matrix, some historical false data should be collected \emph{a priori}, making it less practical and appealing to confront more powerful attackers.

	\emph{Contributions:}
	This paper endeavors to stabilize unknown linear systems in the presence of FDI attacks on the controller-to-actuator channels, through the design of a time-varying state feedback controller from input-state data. 
	A fundamental aspect of this problem revolves around the modeling of unknown FDI attacks in the context of data-driven control. 
	In this regard, we refrain from making assumptions regarding the underlying attack strategy and instead consider a general attack model by only imposing constraints on the switching frequency of attack channels and bounding the attacking power. 
	These considerations enable the reformulation of the FDI attacked system as a switched linear system, incorporating an unknown switching signal and unknown yet bounded subsystems.
	In this context, by exploring the data-based matrix ellipsoid system representation method in \cite{bisoffi2022data}, a data-based set encompassing all switching subsystems is constructed using offline clean data collected from the open-loop healthy system.
	
	To accommodate scenarios where the attacker is too powerful such that stabilizing the set of systems consistent with offline data only becomes infeasible, we seek to merge online input-state observations to tighten the set approximation.
	Subsequently, an online state feedback controller is designed by stabilizing all systems contained in the reduced-size set obtained using both offline clean data and online attacked data.
	These two steps are accomplished by solving two data-based semi-definite programs (SDPs), which allows the controller to adapt to the channel switching of FDI attacks.
	Finally, we establish recursive feasibility for the developed data-based SDPs as well as exponential stability guarantees for the closed-loop data-driven control system under mild conditions on the attack.
	
	In summary, the main contributions of the present paper are as follows.
	
	\begin{itemize}
		\item [c1)]
		A general FDI attack model is presented by constraining the switching frequency and the attack energy, allowing one to characterize the intensity of attacks;
		\item [c2)]
		A data-driven control method is developed, which returns a dynamic feedback controller by solving two data-based SDPs at each time; and,
		\item [c3)]
		Recursive feasibility of the two SDPs as well as exponential stability of the closed-loop attached system is established under mild conditions on the attack.
	\end{itemize}
	
	It is worth pointing out that there are several notable
		differences  with respect to previous works \cite{bisoffi2022data,wu2019optimalswitching}.
	To begin, we modify the SDP in \cite{bisoffi2022data} by updating the data matrices at each time using online input-state data and adapt it to an LQR implementation.
	Due to these differences, stability of the system can be guaranteed only if  recursive feasibility of the proposed SDP is ensured, whose proof deviates considerably from that of \cite{bisoffi2022data}.
	In addition, although \cite{wu2019optimalswitching} has modeled the attack-corrupted system as a switched system too, their focus is on how to design attack strategies to achieve maximum  performance degradation.
	To our best knowledge, no previous work has considered resilient control against FDI attacks when the system model is unknown.

	\emph{Notation:}
	Denote the set of all real numbers, integers, non-negative, and positive integers by $\mathbb{R}$, $\mathbb{Z}$, $\mathbb{N}$, $\mathbb{N}_{+}$, respectively. 
	For a matrix $M$, its rank is given by ${\rm rank}(M)$; 
	if it has full column (row) rank, its left (right) pseudo-inverse is denoted by $M^\dag$.
	Given a vector $x\in \mathbb{R}^{n_x}$, let $\Vert x\Vert$ denote its Euclidean norm.
	The spectral norm of $M$ is given by $\Vert M\Vert$.
	Moreover, $M \succ 0$ ($M \succeq 0$) means $M$ is positive (semi-)definite, and $M \prec 0$ ($M \preceq 0$) means $M$ is negative (semi-)definite.
	For matrices $A$, $B$ and $C$ with compatible dimensions, we abbreviate $ABC(AB)'$ to $AB \cdot C[\star]'$.
	Let $\underline{\lambda}_M$ ($\overline{\lambda}_M$) represent the minimum (maximum) singular value of a matrix $M$, and ${\rm Tr}(M)$ denote the trace of the matrix $M$.

	Given a signal $x: \mathbb{N} \rightarrow \mathbb{R}^{n_x}$ and any $T \in \mathbb{N}_+$, let $x_{[0,T-1]} := [x(0)~x(1)~\cdots~x(T-1)]$ denote a stacked window of the signal $x$ in the time interval $[0,T-1]$.
	The definition of persistence of excitation is adapted from \cite{willems2005note}.
	\begin{definition}[\emph{Persistence of excitation}]\label{def:pe}
		Given any $L \in \mathbb{N}_+$, a signal $x_{[0,T-1]} \in \mathbb{R}^{n_x}$ with $T\ge (n_x + 1)L - 1$ is called persistently exciting of order $L$ if ${\rm {rank}}(H_{L}(x_{[0,T-1]})) = n_x L$ where $H_{L}(x_{[0,T-1]}):=\left[
		\begin{matrix}
			x(0)& x(1) & \ldots & x(T-L)\\
			\vdots & \vdots & \ddots & \vdots \\
			x(L - 1) & x(L) & \ldots & x(T-1)
		\end{matrix}
		\right]$.
	\end{definition}
	\section{Preliminaries and Problem Formulation}\label{sec:preliminaries}
	
	\subsection{Healthy System}\label{sec:preliminaries:sys} 
	Consider the discrete-time linear state feedback system
	\begin{subequations}\label{eq:sys:ideal}
		\begin{align}
			x(t + 1) &= A_{\rm tr}x(t) + B_{\rm tr}u(t)\label{eq:sys:ideal:x}\\ 
			u(t) &= Kx(t)\label{eq:sys:ideal:u}
		\end{align}
	\end{subequations}
	where $x(t)\in \mathbb{R}^{n_x}$ is the state,  $u(t)\in \mathbb{R}^{n_u}$ is the control input, and  $A_{\rm tr}\in\mathbb{R}^{n_x\times n_x}$ and $B_{\rm tr}\in\mathbb{R}^{n_x\times n_u}$ are fixed system matrices. 
	The state feedback controller gain matrix $K$ is to be designed. In this paper, we make the following standing assumptions.
	\begin{assumption}[{Controllability}]\label{as:1:co}
		The pair $(A_{\rm tr}, B_{\rm tr})$ in \eqref{eq:sys:ideal} is unknown but  controllable.
	\end{assumption}
	\begin{assumption}[Offline data]
		\label{as:2:tre}
		For some given integer $T\in\mathbb{N}_+$, input-state data $(u_{[0,T - 1]},x_{[0,T]})$ obtained from the open-loop healthy system \eqref{eq:sys:ideal:x}, are available, where the control input signal $u_{[0,T - 1]}$ is persistently exciting of order $n_x + 1$.
	\end{assumption}

	Assumption \ref{as:2:tre} is very common in the field of data-driven control, as seen in \cite{persis2020data,bisoffi2022data,rotulo2021online,kang2023minimum}. 
	To facilitate the subsequent design and analysis in this paper, we introduce the following data matrices associated with the signals $u_{[0,T - 1]}$ and $x_{[0,T]}$:
	{\setlength{\abovedisplayskip}{5pt}
	\setlength{\belowdisplayskip}{5pt}
	\begin{subequations}\label{eq:X0}
		\begin{align}
			&U_{0} \!=\! [u(0)\cdots u(T \!-\! 1)],~X_{0} \!=\! [x(0)~\cdots~x(T - 1)]\\
			&X_{1} = [x(1)~\cdots~x(T)],~ W_0 = \left[X_0^{\prime}~U_0^{\prime}\right]^{\prime}.
		\end{align}
	\end{subequations}}
	
	Under Assumption \ref{as:2:tre}, it follows from \cite[Corollary 2]{willems2005note} that the rank of $W_0$ satisfies ${\rm rank}(W_0) = n_x + n_u$. This condition ensures that any length-$T$ input-state trajectory of the open-loop system \eqref{eq:sys:ideal:x} can be expressed as a linear combination of the columns of matrix $W_0$; that is, $W_0$ contains complete information about the system's dynamics in the sense that the system can be uniquely represented by using these offline data \cite[Theorem 1]{persis2020data}, i.e., $Z_{\rm tr} := [A_{\rm tr}~B_{\rm tr}]' = (X_1 W_0^{\dag})'$.

	During online operation, we assume that the control input is transmitted over a vulnerable communication channel, e.g. a wireless channel, which is susceptible to FDI attacks. 
	As a result, the integrity of the control input can be compromised, leading to adversarial changes in the system's dynamics \eqref{eq:sys:ideal:x}. 
	In this context, our objective is to design a stabilizing controller in the form of \eqref{eq:sys:ideal:u} to mitigate the impact of FDI attacks and ensure the closed-loop stability of the resultant FDI-corrupted system.

	\subsection{Switched FDI Attack}\label{sec:preliminaries:fdi}
	Before proceeding, we provide an overview of the FDI attack, which are aimed at manipulating the control signal and corrupting the dynamics to degrade system's performance or even cause instability. These attacks are typically modeled as additive inputs on the system state or sensor measurements \cite{pasqualetti2013attack}. Significant efforts have been devoted to designing strategies for launching FDI attacks to induce instability in the healthy system while evading detection \cite{fawzi2014secure,wu2020optimal}. In this paper, our focus is on designing a stabilizing controller for linear systems whose controller-to-actuator channels are subject to FDI attacks, rather than pursuing specific attacking strategies.
	
	We consider a general FDI attack model as in \cite{wu2020optimal}. 
	In the offline phase, the attacker selects a set of attack matrices, one for each representing a possible combination of the controller-to-actuator channels. 
	It further constructs an attack strategy determining the switching times between these combinations and the injected signal.
	During online operation, the attacker eavesdrops on the system state and determines the attack matrix according to its attack strategy. By multiplying the selected attack matrix with the current state, false data can be formed and injected into the corresponding controller-to-actuator channels. Different attack strategies and matrices could lead to different performances. We do not make any specific assumptions about the switching strategy of the attacker but simply constrain the magnitude of the attack matrices and the frequency of switching, which can be seen as limiting the capability and resources of the attacker.
	
	To begin with,  we introduce some assumptions regarding the attacker's behavior, which are general in the context of FDI attacks, e.g., \cite{wu2020optimal,murguia2020security}. To distinguish the offline clean data $(x(t), u(t))$  in \eqref{eq:sys:ideal} from the attack-corrupted data, we denote the online corrupted state and input data as $(x_p(t),u_p(t))$. The following assumption is made.

	\begin{assumption}[FDI attack]
		\label{as:3:FDI}
		The FDI attack satisfies the following conditions.
		\begin{itemize}
			\item [i)]
			The attacker has access to the offline input-state data in Assumption \ref{as:2:tre}. Based on the data, the attacker determines its attack strategy.
			\item [ii)]
			When an FDI attack occurs, the attacker eavesdrops on the plant-to-controller channel to obtain the current system state $x_p(t)$. 
			\item [iii)]
			Based on $x_p(t)$ and the attack strategy, the attacker computes and injects the false data $u_a(t)$ into the state feedback control signal $u_o(t)$, so the actuators implement $	u_p(t) =u_o(t) + u_a(t)$.
		\end{itemize}
	\end{assumption}

	{\setlength{\textfloatsep}{5pt}
	\begin{figure}[t]
		\centering
		\includegraphics[width=8cm]{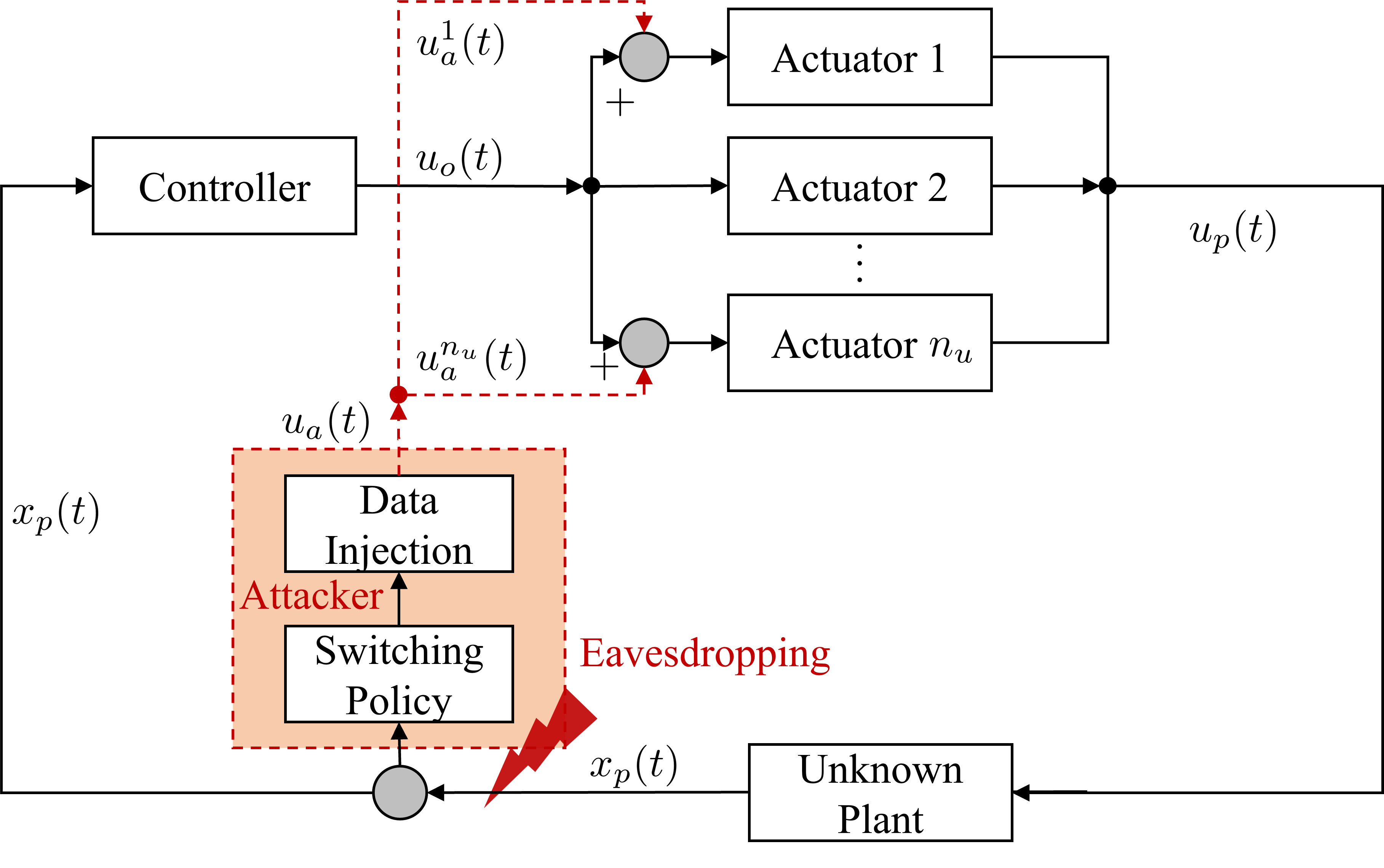}
		\caption{System \eqref{eq:sys} under  FDI attacks on actuators.}
		\label{fig:sys}
		\centering
	\end{figure}}
	
	See Fig. \ref{fig:sys} for a pictorial description of the considered setup. Under Assumption \ref{as:3:FDI},	the FDI attacked system is given by the following equations
	{\setlength{\abovedisplayskip}{5pt}
	\setlength{\belowdisplayskip}{5pt}
	\begin{subequations}\label{eq:sys:c}
		\begin{align}
			x_p(t + 1) &= A_{\rm tr}x_p(t) + B_{\rm tr}u_p(t)\label{eq:sys:c:x}\\ 
			u_p(t) &= u_o(t) +  u_a(t)\label{eq:sys:c:u}\\
			u_o(t) &= K x_p(t).
		\end{align}
	\end{subequations}}

	We consider a strong attacker which has access to and can compromise all controller-to-actuator channels at a time. Specifically, at each time $t$, the attacker may select a subset of these channels based on its strategy from the set $\mathcal{M} := \{1, 2, \ldots, M\}$. 
	Here, $M = \sum_{i = 0}^{n_u} C_{n_u}^i$ represents the total number of attack choices, determined by the combination calculator $C$ as $C_{n_u}^i = n_u!/(i!(n_u-i)!)$, totaling the number of combinations of channels. To model the switching of compromised channels by the attacker across time, we introduce a switching signal $\sigma(t): \mathbb{N} \rightarrow \mathcal{M}$, which is a piece-wise constant function of time taking values in $\mathcal{M}$. The attacker utilizes different gain matrices selected from a prescribed set to corrupt the channels. At each time $t$, the false data are precisely obtained by multiplying the current state $x_p(t)$ with the matrix $D_a^{\sigma(t)}K_a^{\sigma(t)}$, where $D_a^{\sigma(t)}$ is the channel-selection matrix from $\{D_a^j \in \mathbb{R}^{n_u\times n_u}: j \in \mathcal{M}\}$ and $K_a^{\sigma(t)}$ is the attack matrix from $\{K_a^j \in \mathbb{R}^{n_u\times n_x}: j \in \mathcal{M}\}$\footnote{Here, a unique attack matrix $K_a^j$ for each channel-selection $D_a^j$ is considered only for
			illustration of the key idea as well as  ease of computing the total number of switching modes.}. 
	
	\begin{example}
		Let us consider a system as in \eqref{eq:sys:ideal} with $n_u = 3$ actuator channels, resulting in $M = \sum_{i = 0}^{3} C_{3}^i = 8$ different channel combinations $\emptyset, \{1\}, \{2\}, \{1, 2\}, \{1, \\3\}, \{2, 3\}, \{1, 2, 3\}$, which are sequentially indexed by the elements in $\mathcal{M}=\{1, 2, 3, 4, 5, 6, 7, 8\}$. Suppose that at time $t$ the first and third channels (which therefore corresponds to the $6$-th channel combination $\{1,3\}$) are attacked, i.e., $\sigma(t) = 6$, and the associated channel-selection matrix $D_a^{6}$ is given by $D_a^{6}
		=\left[\begin{matrix}
			1&0&0\\
			0&0&0\\
			0&0&1
		\end{matrix}\right].$
	\end{example}

	
	\subsection{System Remodeling and Problem Formulation}
	In the described setup above, the compromised system \eqref{eq:sys:c} under such FDI attacks can be equivalently reformulated as a switched linear system comprising $M$ subsystems, operating under the switching signal $\sigma(t)$, i.e., 
	{\setlength{\abovedisplayskip}{5pt}
	\setlength{\belowdisplayskip}{5pt}
	\begin{subequations}\label{eq:sys:s}
		\begin{align}
			x_p(t + 1) &= A_{\sigma(t)}x_p(t) + B_{\rm tr}u_o(t)\label{eq:sys:s:x}\\
			u_o(t) &= Kx_p(t)\label{eq:sys:s:u}
		\end{align}
	\end{subequations}}
	where $A_{\sigma(t)} := A_{\rm tr} + B_{\rm tr} D_a^{\sigma(t)}K_a^{\sigma(t)}\in\{A_{\rm tr}+B_{\rm tr}D_a^jK_a^j: j\in\mathcal{M} \}$. We use $t_s$ to denote the time when the $s$-th attack occurs, defined as $t_{s} := \min\{t > t_{s - 1}:\sigma(t) \ne \sigma(t_{s - 1}) \}$ with $s \in \mathbb{N}_+$. Without loss of generality, we assume $t_0=0$. Suppose that the system is in mode $j$ at time $t_s$, that is, $\sigma(t) = j$ for all $t \in [t_s, t_{s + 1} - 1]$. In general, the attacker has limited resources. To reflect this fact, we make the following assumptions.
	
	\begin{assumption}[Switching frequency]\label{as:dwell}
		For positive integers $t_1 \le t_2 $, let $N_{\sigma}(t_1, t_2)$ denote the number of discontinuities in signal $\sigma$ over the interval $[t_1, t_2)$. There exist constants $\upsilon \ge 0$ and $\tau \ge 2$, referred to as the chatter bound and the average dwell-time, respectively, which satisfy $N_\sigma(t_1,t_2) \le \upsilon + (t_2 - t_1)/\tau$.
	\end{assumption}
	
	\begin{assumption}[Attacking power] \label{as:4:delta}
		There exists a constant $\phi >0$ such that $\|D_a^{j}K_a^{j}\| \le \phi$ holds for $j \in \mathcal{M}$.
	\end{assumption}
	
	\begin{assumption}[Attacked system]\label{as:5:ctrl}
		The pairs $(A_j, B_{\rm tr})$ for all $j \in \mathcal{M}$ are unknown to both the defender and attacker, but they are each assumed to be controllable.
	\end{assumption}
	
	These assumptions impose restrictions on the switching behavior, attacking power, and knowledge of the attacker, providing a framework for analyzing the cyber-security of linear systems.
	In addition, note that $C_{n_u}^0 = 1$ indicates that no attacks occurs.
	This further means that Assumption \ref{as:1:co} is implied in Assumption \ref{as:5:ctrl}.

	Define $\delta := \phi\Vert  B_{\rm tr}\Vert $.
	Under Assumptions \ref{as:4:delta} and \ref{as:5:ctrl}, it can be easily seen that all switching subsystems $Z_j := [A_j~B_{\rm tr}]'$, $\forall j \in \mathcal{M}$, are contained in a set which are at most $\delta$-far from the unknown true system $Z_{\rm tr} := [A_{\rm tr}~B_{\rm tr}]'$, defined by $\mathcal{B}^{\delta} :=\!\left \{Z = [A~B]': \Vert Z - Z_{\rm tr} \Vert  \le \delta\right\}$.
	Building on this observation, a possible approach to stabilizing the unknown system \eqref{eq:sys:ideal} under the described FDI attacks
	is to design a static feedback controller $K$ that stabilizes all subsystems $(A_j, B_{\rm tr})$, $\forall j \in \mathcal{M}$. Indeed, the
	previous work \cite{wu2019optimalswitching} achieved this by assuming that the injected false data $u_a(t)$ can be collected offline and the matrices $A_j$ can be identified by using a least-squares algorithm. In practice however, obtaining access to the injection signal of the attacker is unrealistic, and finding a common static controller $K$ to stabilize all subsystems is challenging or even impossible, especially when the matrices $\{(A_j, B_{\rm tr})\}_{j\in\mathcal{M}}$ are not known.
	
	To address this issue, we propose designing a dynamic feedback controller $u_o(t) = K(t)x_p(t)$ with the gain matrix $K(t)$ updated on-the-fly. Therefore, the switched system \eqref{eq:sys:s} becomes
	\begin{subequations}\label{eq:sys}
		\begin{align}
			x_p(t + 1) &= A_{\sigma(t)}x_p(t) + B_{\rm tr} u_o(t)\label{eq:sys:x}\\
			u_o(t) &= K(t)x_p(t)\label{eq:sys:u}
		\end{align}
	\end{subequations}
	where the switching signal $\sigma(t):\mathbb{N}\to\mathcal{M}$ is dictated by the attacker and assumed unknown.
	
	Building upon these  preliminaries, we formally present the problem of interest as follows.
	\begin{problem}\label{pro:2}
		Under Assumptions \ref{as:2:tre}---\ref{as:5:ctrl}, design the time-varying state feedback controller $K(t)$ such that the switched system \eqref{eq:sys} achieves exponential stability.
	\end{problem}
	
	The focus of the remaining sections will be on addressing Problem \ref{pro:2}.

	\section{Data-driven Control Against FDI}
	
	\subsection{Data-driven Control Strategy}\label{sec:ctrl:ctrl}
	Learning a data-driven controller to achieve exponential stability of the unknown switched system \eqref{eq:sys} is challenging due to the presence of unknown subsystem matrices, switching signals, and switching times. 
	The fresh idea we advocate here is to learn a controller $K(t)$ stabilizing a minimal subset of systems that contain the active switched subsystem $(A_{\sigma(t-1)},B_{\rm tr})$ and are consistent with the data at time $t$.
	To this aim, we first reformulate the set $\mathcal{B}^{\delta}$ using the offline data.
	In addition, we demonstrate that incorporating online data can help reduce the size of the set of feasible subsystems, increasing the possibility of finding a dynamic stabilizing controller. Finally, by solving two data-based SDPs at each time, we can construct a dynamic controller with guaranteed recursive feasibility for SDP as well as closed-loop system stability.

	Recalling
	$Z_{\rm tr} = [A_{\rm tr}~B_{\rm tr}]' = (X_1 W_0^{\dag})'$,
	the set $\mathcal{B}^{\delta}$ can be equivalently converted into a quadratic matrix inequality (QMI) as follows
	{\setlength{\abovedisplayskip}{5pt}
	\setlength{\belowdisplayskip}{5pt}
	\begin{equation}\label{eq:ellipsoid:delta}
		\mathcal{B}^{\delta} :=\Big \{Z:Z'Z - Z'Z_{\rm tr} -Z_{\rm tr}' Z + \mathbf{C}^{\delta}  \preceq 0\Big\}
	\end{equation}}
	where the constant
	{\setlength{\abovedisplayskip}{5pt}
	\setlength{\belowdisplayskip}{5pt}
	\begin{equation}\label{eq:ABCdelta}
		\mathbf{C}^{\delta} := Z_{\rm tr}'Z_{\rm tr} - \delta^2 I.
	\end{equation}} 
	This set is essentially a matrix ball centered at the true system $Z_{\rm tr}$ with radius $\delta$.
	It is clear that the larger the attacker's power $\delta$ is, the larger the volume of  ball $\mathcal{B}^\delta$ is, and  as a consequence, more ``redundant" systems (those not belonging to the set of subsystems $\{(A_j, B_{\rm tr})\}_{j\in\mathcal{M}}$) this ball contains.
	Although it is possible to search for a static data-driven controller $K$ stabilizing all the systems in $\mathcal{B}^\delta$ using Petersen's lemma \cite{bisoffi2022data} based on the offline data, this can be rather challenging especially when $\delta$ is large.
	In our proposal, we first leverage the online state-input data to obtain a tighter approximation of the set of feasible systems
	on top of $\mathcal{B}^{\delta}$, and subsequently search for a dynamic controller stabilizing all systems in the reduced-size set.
	
	During online operation, we assume that the initial conditions $u_p(0)$ and $x_p(0)$ are arbitrary. 
	At time $t \in \mathbb{N}_+$, we have observed the online input-state data $(x_p(k-1),u_o(k-1),x_p(k))$ for all  $k\le t$. Nonetheless, due to the unknown switching signal $\sigma(\cdot)$, we do not know how exactly these data correspond to the $M$ subsystems since they likely come from multiple subsystems. Therefore, it is difficult to directly employ all these online data to design a controller $K(t)$ to stabilize all subsystems. 
	To bypass this challenge, we propose to combine only the most fresh data $(x_p(t-1),u_o(t-1),x_p(t))$ from the subsystem $(A_{\sigma(t-1)},B_{\rm tr})$ and the set of offline data from the healthy system $(A_{\rm tr}, B_{\rm tr})$ to design the controller $K(t)$ on the fly. 
	
	To this end, let us consider the switched system \eqref{eq:sys} at any time $t \in \mathbb{N}_+$, for which we have observed the online data $(x_p(t-1),u_o(t-1),x_p(t))$. The set of matrices ${Z}_t = [{A}_t~{B}_t]'$ that can generate $(x_p(t-1),u_o(t-1),x_p(t))$ is given by 
	{\setlength{\abovedisplayskip}{5pt}
	\setlength{\belowdisplayskip}{5pt}
	\begin{align}
		\mathcal{E}_{t} \!=\! \bigg\{\!Z_t :& \  \forall d(t-1) \in \mathbb{R}^{n_x}~{\rm such~that}~d(t\!-1)d'(t\!-1) \!=\! 0,\nonumber\\
		&\quad \ x_p(t) = Z_t'\left[\begin{matrix}
			x_p(t-1)\\
			u_o(t-1)
		\end{matrix}\right] + d(t-1)\bigg\}.\label{eq:epsilont}
	\end{align}}
	Since $d(t-1)d'(t-1) = 0$ in \eqref{eq:epsilont} can be implied by $d(t-1)d'(t-1)\preceq 0$, the result in \cite[(12)]{van2021finsler} has demonstrated its equivalence to the following QMI
	{\setlength{\abovedisplayskip}{5pt}
	\setlength{\belowdisplayskip}{5pt}
	\begin{equation}
		[I\ \ d(t-1)]\left[
		\begin{matrix}
			0&0\\
			0&I
		\end{matrix}
		\right]\left[\begin{matrix}
			I\\
			d'(t-1)
		\end{matrix}\right] \preceq 0.\label{eq:qmi}
	\end{equation}} 
	Moreover, we can rewrite $d(t-1)=x_p(t ) - Z_t'\left[\begin{matrix}
		x_p(t-1)\\
		u_o(t-1)
	\end{matrix}\right]$ and plug it into the QMI in \eqref{eq:qmi}, the set $\mathcal{E}_t$ boils down to	
	{\setlength{\abovedisplayskip}{5pt}
	\setlength{\belowdisplayskip}{5pt}
	\begin{equation*}
		\mathcal{E}_{t} = \left\{Z_t :
		\left[
		I\ \  Z_t'
		\right]
		\left[
		\begin{matrix}
			I&x_p(t)\\
			0&-x_p(t-1)\\
			0&-u_o(t-1)
		\end{matrix}
		\right]
		\cdot
		\left[
		\begin{matrix}
			0&0\\
			0&I
		\end{matrix}
		\right][\star]'\preceq 0\right\} .
	\end{equation*}}
	Upon expanding the QMI and introducing	
	{\setlength{\abovedisplayskip}{5pt}
	\setlength{\belowdisplayskip}{5pt}
	\begin{subequations}\label{eq:ellipsoid:t:abc}
		\begin{align}
			&\mathbf{A}_{t} := 
			\left[
			\begin{matrix}
				x_p(t -1)\\
				u_o(t-1 )
			\end{matrix}
			\right]
			\left[
			\begin{matrix}
				x_p(t -1)\\
				u_o(t -1)
			\end{matrix}
			\right]'\label{eq:ellipsoid:t:a}\\
			&\mathbf{B}_{t}: = -\left[
			\begin{matrix}
				x_p(t -1)\\
				u_o(t -1)
			\end{matrix}
			\right]x_p'(t),~\mathbf{C}_{t} := x_p(t)x_p'(t)\label{eq:ellipsoid:t:bc}
		\end{align}
	\end{subequations}}
	one can further re-express the set $\mathcal{E}_t$ as follows
	{\setlength{\abovedisplayskip}{5pt}
	\setlength{\belowdisplayskip}{5pt}
	\begin{equation}
		\mathcal{E}_t = \Big\{Z_t : Z_t' \mathbf{A}_{t}Z_t + Z_t' \mathbf{{B}}_{t} + \mathbf{{B}}_{t}' Z_t + \mathbf{{C}}_{t} \preceq 0\Big\}\label{eq:ellipsoid:t}
	\end{equation}}
	which is a matrix ellipsoid too. 
	It is evident that  the active subsystem $[A_{\sigma(t-1)} ~B_{\rm tr}]' \in \mathcal{E}_t$.

	It is worth noting that the ellipsoid $\mathcal{E}_t$ is degenerate and unbounded, even in the simplest case when $n_u = n_x = 1$  \cite[Lemma 2]{bisoffi2022data}, due to the rank deficiency of matrix $\mathbf{A}_{t}$. Therefore, one cannot directly capitalize on Petersen's lemma \cite{bisoffi2022data} to design the controller $K(t)$. 

	Note from  \eqref{eq:ellipsoid:delta} that all subsystems are contained in $\mathcal{B}^{\delta}$, which implies $Z_{\sigma(t-1)} = [A_{\sigma(t-1)}~B_{\rm tr}]' \in \mathcal{B}^{\delta}$ for all $t$. Combining this with \eqref{eq:ellipsoid:t}, we conclude that
	{\setlength{\abovedisplayskip}{5pt}
	\setlength{\belowdisplayskip}{5pt}
	\begin{equation}\label{eq:set:int}
		Z_{\sigma(t-1)} \in \mathcal{E}_{t} \cap \mathcal{B}^{\delta},\quad \forall t \in \mathbb{N}_+.
	\end{equation} }
\begin{remark}[$Z_t$ vs. $Z_{\sigma(t-1)}$]
	Symbols $Z_t$ and $Z_{\sigma(t-1)}$ are used to refer to different types of matrices.
	Specifically, $Z_{\sigma(t-1)}$  represents the actual attacked system at time $t-1$.
	On the other hand, $Z_t$ represents any matrices within the set $\mathcal{E}_t$, which include not only the actual attacked system matrices $Z_{\sigma(t-1)}$, but also other matrices consistent with the data $x_p(t)$, $x_p(t-1)$,  $u_p(t-1)$.
\end{remark}
	The problem thus reduces to finding a controller $K(t)$ to stabilize all the systems in the set $\mathcal{E}_{t} \cap \mathcal{B}^{\delta}$ for all $t$.
	Nonetheless, it is difficult to analytically express the set $\mathcal{E}_{t} \cap \mathcal{B}^{\delta}$. 
	In \cite[Section 4.4]{bisoffi2022data}, a method for computing a set that over-approximates $\mathcal{E}_{t} \cap \mathcal{B}^{\delta}$ while minimizing the volume of the set is presented. Before moving on, we provide a brief review of this method. 
	
	Consider the following set
	{\setlength{\abovedisplayskip}{5pt}
	\setlength{\belowdisplayskip}{5pt}
	\begin{equation*}
		\mathcal{I}_{t} := \Big\{Z_t: Z_t' \bar{\mathbf{A}}_tZ_t + Z_t' \bar{\mathbf{B}}_{t} + \bar{\mathbf{B}}_t'Z_t + \bar{\mathbf{C}}_t \preceq 0\Big\}
	\end{equation*}}
	with  $\bar{\mathbf{A}}_{t} = \bar{\mathbf{A}}_{t}' \succ 0$ and $\bar{\mathbf{C}}_{t} = \bar{\mathbf{B}}_{t}' \bar{\mathbf{A}}_{t}^{-1}\bar{\mathbf{B}}_{t} - I$. 
	This set can be equivalently rewritten as follows 
	{\setlength{\abovedisplayskip}{5pt}
	\setlength{\belowdisplayskip}{5pt}
	\begin{equation*}
		\mathcal{I}_{t} = \Big\{Z_t: (Z_t + \bar{\mathbf{A}}_t^{-1}\bar{\mathbf{B}}_t)' \bar{\mathbf{A}}_t(Z_t + \bar{\mathbf{A}}_t^{-1}\bar{\mathbf{B}}_t) -I \preceq 0\Big\}
	\end{equation*}}
	which represents a matrix ellipsoid centered at $-\bar{\mathbf{A}}_t^{-1}\bar{\mathbf{B}}_t$ with volume scaling with $-\log \det(\bar{\mathbf{A}}_t)$ as discussed in \cite[Section 4.4]{bisoffi2022data}.
	Recalling the matrices $\mathbf{C}^{\delta}$ in \eqref{eq:ABCdelta} and  $(\mathbf{A}_{t},\mathbf{B}_{t},\mathbf{C}_{t})$ in \eqref{eq:ellipsoid:t:abc}, let $\bar{\mathbf{A}}^\ast_{t}$ and $\bar{\mathbf{B}}^\ast_t$ denote the optimal solutions of the following SDP problem \cite[(34)]{{bisoffi2022data}}
	\begin{subequations}\label{eq:intersection}
		\begin{align}
			\min_{\mathbf{\bar{A}}_t, \mathbf{\bar{B}}_t, \atop \tau_1 \ge 0, \tau_2 \ge 0} &-\log \det(\mathbf{\bar{A}}_t) \\
			{\rm s.t.}~~~& \mathbf{\bar{A}}_t = \mathbf{\bar{A}}_t' \succ 0\\
			&\left[
			\begin{matrix}
				-I &\mathbf{\bar{B}}_t'  & \mathbf{\bar{B}}_t' \\
				\mathbf{\bar{B}}_t & \mathbf{\bar{A}}_t & 0\\
				\mathbf{\bar{B}}_t &0& -\mathbf{\bar{A}}_t
			\end{matrix}
			\right] 
			- \tau_1\left[
			\begin{matrix}
				\mathbf{C}^{\delta}  & -Z_{\rm tr}' & 0\\
				-Z_{\rm tr} & I & 0\\
				0 &0 & 0
			\end{matrix}
			\right] \nonumber\\
			&-\tau_2\left[
			\begin{matrix}
				\mathbf{C}_{t}  & \mathbf{B}_{t}^{\prime} & 0\\
				\mathbf{B}_{t} & \mathbf{A}_{t} & 0\\
				0 & 0 & 0
			\end{matrix}
			\right]
			\preceq 0.
		\end{align}
	\end{subequations}
	
	Letting $\bar{\mathbf{C}}_t^\ast = \bar{\mathbf{B}}_t^{\ast\prime}\bar{\mathbf{A}}_t^{\ast-1}\bar{\mathbf{B}}^\ast_t - I$, the minimum-volume matrix ellipsoid containing the set $\mathcal{E}_{t} \cap \mathcal{B}^{\delta}$ is given by
	\begin{align}
		\mathcal{I}_{t}^\ast&= \Big\{Z_t: Z_t' \bar{\mathbf{A}}_t^\ast Z_t + Z_t' \bar{\mathbf{B}}_{t}^\ast + \bar{\mathbf{B}}_t^{\ast\prime}Z_t + \bar{\mathbf{C}}_t^\ast \preceq 0\Big\}\label{eq:elipsoid:inter}	 \\
		&  = \Big\{Z_t: (Z_t \!+\! \bar{\mathbf{A}}_t^{\ast-1}\bar{\mathbf{B}}^\ast_t)' \cdot \bar{\mathbf{A}}^\ast_t[\star]' -I \preceq 0\Big\}\label{eq:elipsoid:inter:eq}.
	\end{align}
	In addition, since the objective function $-\log \det(\mathbf{\bar{A}}_t)$ is a convex function, the SDP problem \eqref{eq:intersection} with linear constraints is convex, and it can be solved efficiently using off-the-shelf convex optimization toolboxes.
	
\begin{figure}[t]
		\centering
		\includegraphics[width=8cm]{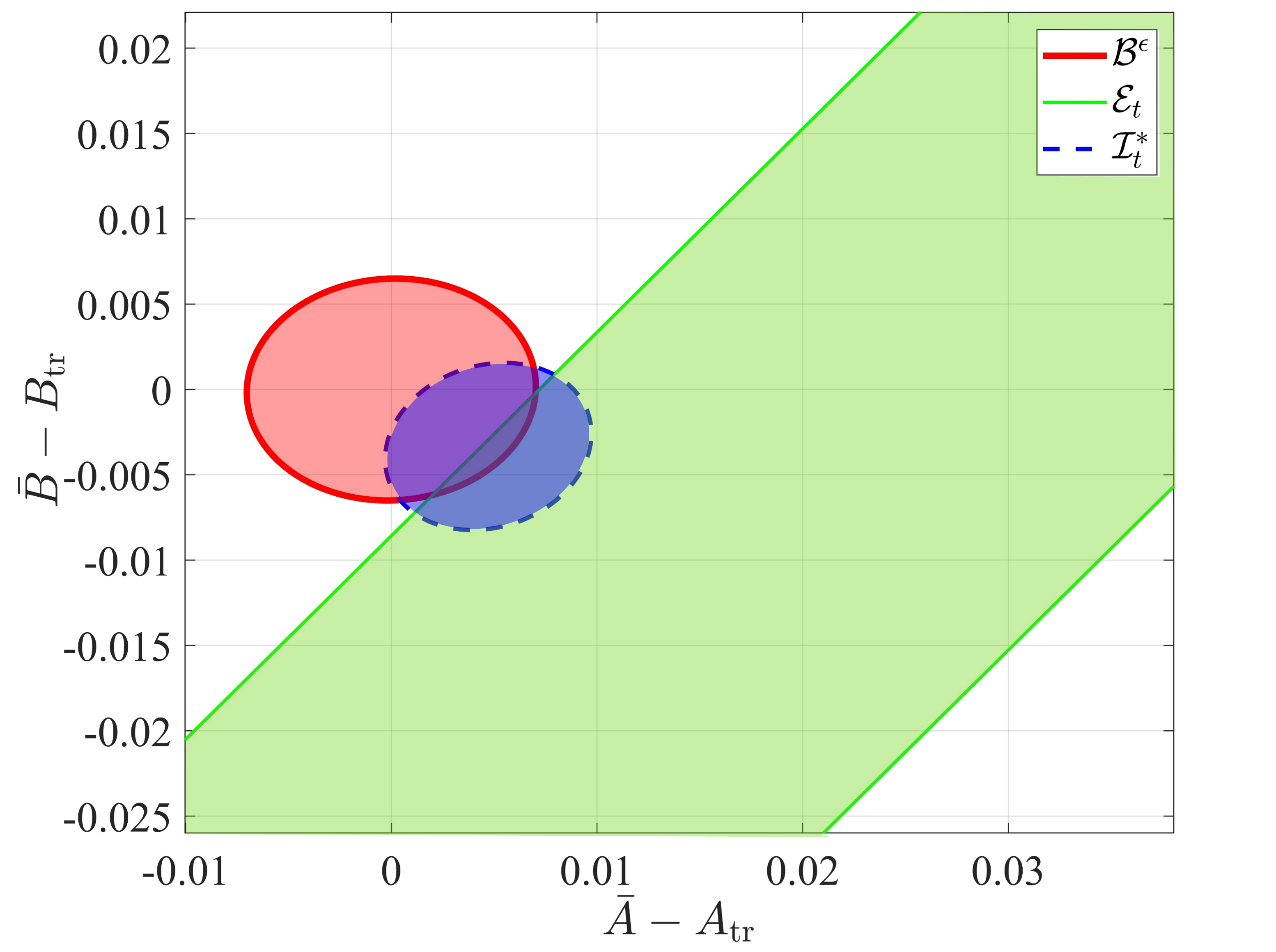}\\
		\caption{Sets $\mathcal{B}^{\delta}$ (red solid circle), $\mathcal{E}_t$ (green solid ellipsoid), and $\mathcal{I}^*_t$ (blue dashed-line ellipsoid) for a first-order system.\label{fig:intersection}}
		\centering
	\end{figure}

	\begin{example}
		To provide a better understanding, the relationships between sets $\mathcal{B}^{\delta}$, $\mathcal{E}_t$, and $\mathcal{I}_t^\ast$ are illustrated in Fig. \ref{fig:intersection} using a first-order system with $A_{\rm tr} \!=\! 1.021$, $B_{\rm tr} \!=\! 0.041$, and $\delta \!=\! 0.025$.
	\end{example}

	From the figure,  it is evident that if we can find a matrix $K(t)$ that stabilizes all the systems
	$Z_t =[A_t~B_t]' \in \mathcal{I}_t^\ast$, then we can stabilize the system $(A_{\sigma(t-1)},B) \in \mathcal{I}_t^\ast$ using $K(t)$. 
	Finding a stabilizing controller for a set of systems characterized by a QMI has been studied in \cite{bisoffi2022data}, in which such a controller $K(t)$ was obtained by resorting to Petersen's lemma. 
	Building on this idea and inspired by the formulations for data-driven LQR and LQG control recently in \cite{depersis2021lowcomplexity,liu2023learning}, we formulate the following SDP parameterized by the coefficient matrices defining the QMI \eqref{eq:elipsoid:inter} by balancing the system's performance and the robustness against FDI attacks

		\begin{subequations}\label{eq:lqr:on}
		\begin{align}
			\min_{\gamma , P, Y, L,Q} & \ \gamma \nonumber\\
			{\rm s.t.}~~~& \left[
			\begin{matrix}
				-\epsilon_1 P - \bar{\mathbf{C}}_t^\ast & 0& \bar{\mathbf{B}}_t^{\ast\prime} \\
				0 & -P & [P~ Y'] \\
				\bar{\mathbf{B}}_t^{\ast}& [P~ Y']' & -\bar{\mathbf{A}}_t^\ast\\
			\end{matrix}
			\right]\preceq 0\label{eq:lqr:on:1}\\
			&P  \succ 0\label{eq:lqr:on:2}\\
			&\left[
			\begin{matrix}
				L & Y\\
				Y' & P
			\end{matrix}
			\right] \succeq 0\label{eq:lqr:on:3}\\
			&\left[
			\begin{matrix}
				Q & I\\
				I & P
			\end{matrix}
			\right] \succeq 0\label{eq:lqr:on:4}\\
			&{\rm Tr}(P) + {\rm Tr}(L) + \epsilon_2 \Vert Q\Vert\le \gamma \label{eq:lqr:on:5}.
		\end{align}
	\end{subequations}
	Here, Constrains \eqref{eq:lqr:on:1} and \eqref{eq:lqr:on:2} guarantee that the resultant gain matrix $K(t)$ constructed by the optimal solutions is stabilizing; Constrains \eqref{eq:lqr:on:3}--\eqref{eq:lqr:on:5} are adapted from the robust data-driven LQR ensuring optimal performance of the consequent $K(t)$; and, the parameters $\epsilon_1 \in (0,1)$ and $\epsilon_2 >0$ are arbitrary, which balance the system's performance as well as feasibility of the SDP.
	Assume for now the SDP is feasible at each time $t$, and let $(\gamma^*(t), P^*(t), Y^*(t), L^*(t),Q^\ast(t))$ denote any optimal solution. The stabilizing controller $u_o(t) = K(t)x_p(t)$ for all systems in $\mathcal{I}_t^\ast$ 
	can be designed as follows
	\begin{equation}\label{eq:Kk}
		K(t) = {Y}^*(t) ({P}^{*}(t))^{-1}.
	\end{equation}
	
	Based on \eqref{eq:intersection}--\eqref{eq:Kk}, the proposed data-driven controller is summarized in Algorithm \ref{alg:ctrl}.
	However, successful implementation of Algorithm \ref{alg:ctrl} hinges on the feasibility of SDP \eqref{eq:lqr:on}.
	Moreover, even if its feasibility is always ensured, it can be observed that at the switching time $t = t_s$, there is a mismatch between the mode of the controller and that of the system; that is, the active system is now $(A_{\sigma(t_s)}, B_{\rm tr})$ while the computed controller $K(t)$ is designed for stabilizing the system $(A_{\sigma(t_{s-1})}, B_{\rm tr})$.
	This mismatch may lead to the divergence of state trajectory if it happens frequently.
	To address problem \ref{pro:2} using Algorithm \ref{alg:ctrl}, we encounter two difficulties, which will be tackled in the following subsection, i.e., d1) how to ensure the feasibility of SDP \eqref{eq:lqr:on} for all $t \in \mathbb{N}_+$? and, d2) under what conditions, can we stabilize the system \eqref{eq:sys}?

	
	\begin{algorithm}[t]
		\caption{Data-driven control against FDI.}
		\label{alg:ctrl}
		\begin{algorithmic}[1]
			\small
			\State {\bfseries Offline:} 
			Collect input-state data $(u_{[0,T\!-1]}, x_{[0,T]})$ and
			form matrices $X_0,X_1\!,U_0,W_0$ as in \eqref{eq:X0}.
			Compute $\mathcal{B}^{\delta}$ as in \eqref{eq:ellipsoid:delta}.
			\State {\bfseries Online:} 
			Give initial conditions $x_p(0)$, $u_o(0)$ and 
			constants $\epsilon_1 \in (0,1)$, $\epsilon_2 >0$. 
			\State For $t =1,2,3,\ldots$, do
			\begin{itemize}
				\item[1)]
				Compute the matrix ellipsoid $\mathcal{E}_t$ in \eqref{eq:ellipsoid:t} based on $u_o(t - 1)$, $x_p(t - 1)$, and $x_p(t)$.
				\item[2)]
				Solve the SDP in \eqref{eq:intersection} to obtain an over-approximation $\mathcal{I}_t^\ast$ of the intersection  $\mathcal{B}^{\delta} \cap \mathcal{E}_t$ as in \eqref{eq:elipsoid:inter}.
				\item[3)]
				Solve the SDP in \eqref{eq:lqr:on} for the set $\mathcal{I}_t^\ast$ and denote the solution as $(\gamma^*(t), P^*(t), Y^*(t), L^*(t),Q^*(t))$.
				\item[4)]
				Compute the control input $u_o(t) = K(t)x_p(t)$, where $K(t) = Y^*(t)(P^*(t))^{-1}$ is given in \eqref{eq:Kk}.
				\item[5)]
				Set $t = t+1$ and go back to 1).
			\end{itemize}
		\end{algorithmic}
	\end{algorithm}
	
	\subsection{Theoretical Guarantees}
	
	In this subsection, we focus on answering the two questions in the previous subsection. The following result shows that if the attacking power $\delta$ is small, SDP \eqref{eq:lqr:on} is feasible for all $t \in \mathbb{N}_+$ and the associated controller $K(t)$ is stabilizing.
	
	\begin{theorem}\label{thm:lqr}
		Under Assumptions \ref{as:1:co}---\ref{as:5:ctrl}, let $U_0$, $X_0$ and $X_1$ be given in \eqref{eq:X0}. Consider arbitrary initial conditions $x_p(0)$ and $u_o(0)$ for the system \eqref{eq:sys} with Algorithm \ref{alg:ctrl}. 
		There exist constants $\bar{\delta} >0$ and $\underline{\epsilon}_1 \in (0,1)$ such that for all $\delta \in [0,\bar{\delta})$ and $\epsilon_1 \in (\underline{\epsilon}_1,1)$, SDP \eqref{eq:lqr:on} is feasible for any  $t \in \mathbb{N}_+$ with an optimal solution $(\gamma^*(t), P^*(t), Y^*(t), L^*(t),Q^*(t))$. Let $K(t) = Y^*(t)(P^*(t))^{-1}$. Then the following statements hold true:
		\begin{itemize}
			\item [s1)] Let $j =\sigma(t_s)$ denote the active subsystem for all $t \in [t_s,t_{s+1}-1]$.
			Then the controller $K(t)$ stabilizes the subsystem $(A_j,B_{\rm tr})$. 
			\item [s2)] There exists a constant $\kappa>0$ such that $\Vert K(t)\Vert \le \kappa$ for all $t \in \mathbb{N}_+$.
		\end{itemize} 
	\end{theorem}

		\begin{pf}
		Consider any $s \in \mathbb{N}$ and $t \in [t_s+1,t_{s + 1}]$, and let $j =\sigma(t_s)$ denote the active subsystem $Z_{\sigma(t-1)} = Z_j = [A_j~B_{\rm tr}]'$.  It is clear from the previous subsection that $Z_j  \in \mathcal{E}_{t} \cap \mathcal{B}^{\delta} \subseteq \mathcal{I}_t^\ast$. 
		The key idea of the proof is to construct a candidate solution for 
		SDP \eqref{eq:lqr:on}. 
		To this end, we explore the data-driven LQR formulation of subsystem $(A_j,B_{\rm tr})$ by showing that, under certain conditions on the attacking power $\delta$ and the parameter $\epsilon_1$, a candidate solution of SDP \eqref{eq:lqr:on} can be constructed from the LQR solution.
		This establishes the feasibility of SDP \eqref{eq:lqr:on}, and statements s1) and s2) can be proved.
		
		\emph{i) Proof of Feasibility of SDP \eqref{eq:lqr:on}}
		
		According to \eqref{eq:elipsoid:inter:eq}, although the subsystem $Z_j = [A_{j}~B_{\rm tr}]' \in \mathcal{I}_t^\ast$ is active for all $t \in [t_s + 1, t_{s + 1}]$, the matrix $Z_j$ is generally not the center of the ellipsoid $\mathcal{I}_t^\ast$. 
		This renders the feasibility and stability analysis rather complicated.
		To overcome this difficulty, we construct a matrix ball $\mathcal{B}^{\delta_j}$ centered at $Z_j$ with some radius $\delta_j$ such that  $\mathcal{I}_t^\ast \subseteq \mathcal{B}^{\delta_j}$.
		Hence, to prove the feasibility of SDP \eqref{eq:lqr:on} for $\mathcal{I}_t^\ast$, it suffices to prove that for the ball $\mathcal{B}^{\delta_j}$.
		Building on this observation, we start by defining $\mathcal{B}^{\delta_j}$, and subsequently review the data-driven LQR formulation, based on which we construct a candidate solution of SDP \eqref{eq:lqr:on} for set $\mathcal{B}^{\delta_j}$.
		
		Consider the ball centered at $Z_j$ with radius $\delta_j$ as follows
		\begin{equation}
			\label{eq:elip:on:big}
			\mathcal{B}^{\delta_j} := \Big\{Z: Z'Z	 - Z' Z_j -Z_j'Z+\mathbf{C}^{\delta_j} \preceq 0 \Big\}
		\end{equation}
		where $			\mathbf{C}^{\delta_j} := Z_j'Z_j - \delta_j^2I
		$. 
		Since $\mathcal{I}_t^\ast$ is bounded, there exists some constant $\bar{\delta}_j \in (0, \delta]$ such that for all $\delta_j \in (\bar{\delta}_j,\delta]$, it holds that $ \cup_{t=t_s + 1}^{t = t_{s + 1}} \mathcal{I}_t^\ast \subseteq \mathcal{B}^{\delta_j} $ for any $s \in \mathbb{N}$.
		This indicates that if SDP \eqref{eq:lqr:on} is feasible for set $\mathcal{B}^{\delta_j}$,  it is feasible for $\mathcal{I}_t^\ast$ for all $t \in [t_s + 1, t_{s + 1}]$ too.
		
		By replacing the matrices $\bar{\mathbf{A}}_t^\ast$, $\bar{\mathbf{B}}_t^\ast$ and $\bar{\mathbf{C}}_t^\ast$ in \eqref{eq:lqr:on} with $I$, $-Z_j$, and $\mathbf{C}^{\delta_j}$, respectively, we formulate the following SDP for set $\mathcal{B}^{\delta_j}$ and continue to derive conditions for its feasibility
		\begin{subequations}\label{eq:lqr:ball}
				\begin{align}
					\min_{\gamma , P, Y, L,Q} & \ \gamma \nonumber\\
					{\rm s.t.}~~& \left[
					\begin{matrix}
						-\epsilon_1 P - \mathbf{C}^{\delta_j} & 0& -Z_j'\\
						0 & -P & [P~ Y'] \\
						-Z_j & [P~ Y']' & -I\\
					\end{matrix}
					\right]\preceq 0\label{eq:lqr:ball:1}\\
					&P  \succ 0\label{eq:lqr:ball:2}\\
					&\left[
					\begin{matrix}
						L & Y\\
						Y' & P
					\end{matrix}
					\right] \succeq 0\label{eq:lqr:ball:3}\\
					&\left[
					\begin{matrix}
						Q & I\\
						I & P
					\end{matrix}
					\right] \succeq 0\label{eq:lqr:ball:4}\\
					&{\rm Tr}(P) + {\rm Tr}(L) + \epsilon_2 \Vert Q\Vert\le \gamma \label{eq:lqr:ball:5}.
				\end{align}
		\end{subequations}

		Before proceeding, we recall Petersen's lemma in \cite{bisoffi2022data}, which plays an important role in the rest of the proof.
		Since $\mathcal{B}^{\delta_j}$ is bounded, it has been demonstrated in \cite[Theorem 1]{bisoffi2022data} that  
		the feasibility problem
		\begin{align}
			\label{eq:peter:ori}
			{\rm find}~~&Y,\, P = P' \succ 0\\
			{\rm s.t.\ }~& [A~B]
			\left[
			\begin{matrix}
				P\\
				Y
			\end{matrix}
			\right] \cdot P^{-1}[\star]' - P \prec 0,~~\forall [A~B]'\in \mathcal{B}^{\delta_j} \nonumber 
		\end{align}
		is equivalent to that of			\begin{align}\label{eq:peter:mu}
				{\rm find}~~&Y, \, P = P' \succ 0, \, \lambda >0 \\
				{\rm s.t.\ }~& \left[
				\begin{matrix}
					- \lambda P - \mathbf{C}^{\delta_j}  & 0& -Z_j'\\
					0 & -\lambda P & \lambda[P~ Y'] \\
					-Z_j& \lambda[P~ Y']' & -I\\
				\end{matrix}
				\right] \prec 0.\nonumber
		\end{align}
		
		To show the feasibility SDP \eqref{eq:lqr:ball}, we begin by considering the set $\mathcal{B}^{\delta_j}$ with $\delta_j = 0$.
		In this case, the matrix $\mathbf{C}^{\delta_j}$ reduces to $\mathbf{C}^{\delta_j}=\mathbf{C}^{0} = Z_j'Z_j$.
		Recall from Assumption \ref{as:5:ctrl} that the system $[A_j ~B_{\rm tr}]$ is controllable. Thus, there is a unique LQR controller. Building on \cite[(9)]{depersis2021lowcomplexity}, the LQR controller gain for system $[A_j ~B_{\rm tr}]$ can be found by using
		the optimal solution of the following SDP
		\begin{subequations}
				\label{eq:lqr:peter:ori}
				\begin{align}
					(\tilde{\gamma}_j,~&\!\!\tilde{P}_j, \tilde{Y}_j, \tilde{L}_j):=
					\arg	\min_{\gamma , P, Y, L} \gamma \nonumber\\
					{\rm s.t.}~~& [{A}_j~B_{\rm tr}]
					\!\left[
					\begin{matrix}
						P\\
						Y
					\end{matrix}
					\right]\cdot  P^{-1}[\star]' - P + I \preceq  0 \label{eq:lqr:peter:ori:1}\\
					&P \succeq I \label{eq:lqr:peter:ori:2}\\
					&\left[
					\begin{matrix}
						L & Y\\
						Y' & P
					\end{matrix}
					\right] \succeq 0 \label{eq:lqr:peter:ori:3}\\
					&{\rm Tr}(P) + {\rm Tr}(L)\le \gamma \label{eq:lqr:peter:ori:4}.
				\end{align}
		\end{subequations}

		Moreover, at the optimum, constraint \eqref{eq:lqr:peter:ori:1} becomes
		\begin{equation}\label{eq:py1}
			[A_j~B_{\rm tr}]
			\left[
			\begin{matrix}
				\tilde{P}_{j}\\
				\tilde{Y}_{j}
			\end{matrix}
			\right] \cdot\tilde{P}_{j}^{-1}[\star]' - \tilde{P}_{j} \preceq -I \prec 0
		\end{equation}
		which implies \eqref{eq:peter:ori} with $\delta_j = 0$.
		Since \eqref{eq:peter:ori} is equivalent to \eqref{eq:peter:mu}, there exist some constants $\lambda_j >0$ and ${\rho}_{j} >0$ such that
		\begin{equation}\label{eq:rho}
				\left[
				\begin{matrix}
					- \lambda_j  \tilde{P}_{j} - Z_j'Z_j  & 0& -Z_j'\\
					0 & -\lambda_j \tilde{P}_{j} & \lambda_j[\tilde{P}_{j}~ \tilde{Y}_{j}'] \\
					-Z_j& \lambda_j[\tilde{P}_{j}~ \tilde{Y}_{j}']' & -I\\
				\end{matrix}
				\right] \preceq -{\rho}_{j}I.
		\end{equation}
		
		Based on this inequality, we consider $\delta_j \ge 0$ and derive conditions on $\epsilon_1$ and $\delta_j$ such that a candidate solution of SDP \eqref{eq:lqr:ball} can be constructed using $(\tilde{\gamma}_j, \tilde{P}_j, \tilde{Y}_j, \tilde{L}_j)$. 
		The remaining proof is divided into two parts depending on whether $\lambda_j \in (0,1)$ or $\lambda_j \ge 1$.

		\newcounter{TempEqCnt} 
		\setcounter{TempEqCnt}{\value{equation}} 
		\setcounter{equation}{29} 
		\begin{figure*}[!t]
			\footnotesize
			\begin{align}
					&\left[
					\begin{matrix}
						-\epsilon_1 \mu_j^{\lambda_j} \tilde{P}_{j}^{\lambda_j} \!-\! \mathbf{\bar{C}}^{\delta_j}  \!& 0\!\!\!\!\!& -Z_j'\\
						0 \!& -\mu_j^{\lambda_j}\tilde{P}_{j}^{\lambda} \!& \mu_j^{\lambda_j}[\tilde{P}_{j}^{\lambda_j}~ \tilde{Y}_{j}^{\lambda_j\prime}] \\
						-Z_j\!& \mu_j^{\lambda_j}[\tilde{P}_{j}^{\lambda_j}~ \tilde{Y}_{j}^{\lambda_j\prime}]' \!& -I\\
					\end{matrix}
					\right]  \!{\preceq} \!\left[
					\begin{matrix}
						-\mu_j^{\lambda_j} \rho_{j} I \!+ (\mu_j^{\lambda_j} (1 \!-\! \epsilon_1) \tilde{P}_{j}^{\lambda_j} \!-\! (1-\!\!\mu_j^{\lambda_j})Z_{j}'Z_{j} \! + \!\delta_j^2 I) \!\!\!\!& 0\!\!\!& (\mu_j^{\lambda_j} \!-\! 1) Z_{j}'\\
						0 \!\!\!\!& -\!\mu_j^{\lambda_j} \rho_{j} I \!\!& 0 \\
						(\mu_j^{\lambda_j} \!-\! 1) Z_{j}\!\!\!\!& 0\!\!\! & -\!\mu_j^{\lambda_j} \rho_{j} I+\! (\mu_j^{\lambda_j} \!-\! 1)I\\
					\end{matrix}
					\right] \label{eq:epsilon:delta}
			\end{align}
			\setcounter{equation}{\value{MYtempeqncnt}}
			\hrulefill
			\vspace*{4pt}
		\end{figure*}
		\setcounter{equation}{\value{TempEqCnt}} 
		
		\textbf{Case 1: $\lambda_j \in (0,1)$.}
		
		When $\lambda_j \in (0,1)$, substituting \eqref{eq:rho} into constraint \eqref{eq:lqr:ball:1} yields		\begin{align}\label{eq:mu:rho:2}
				&~\left[
				\begin{matrix}
					-\lambda_{j}\tilde{P}_{j} - Z_{j}'Z_{j} & 0& -Z_{j}'\\
					0 & -\lambda_{j}\tilde{P}_{j} & \lambda_{j}[\tilde{P}_{j}~ \tilde{Y}_{j}'] \\
					-Z_{j}& \lambda_{j}[\tilde{P}_{j}~ \tilde{Y}_{j}']' & -I\\
				\end{matrix}
				\right] \nonumber\\
				&~~~+ \left[
				\begin{matrix}
					(1 - \epsilon_1)\lambda_{j}\tilde{P}_{j}  + \delta_j^2 I & 0& 0\\
					0 & 0 & 0 \\
					0& 0 & 0\\
				\end{matrix}
				\right] \nonumber\\
				&\preceq - {\rho}_{j} I + \left[
				\begin{matrix}
					(1 - \epsilon_1)\lambda_{j}\tilde{P}_{j}  + \delta_j^2 I & 0& 0\\
					0 & 0 & 0 \\
					0& 0 & 0\\
				\end{matrix}
				\right]. 
		\end{align}
		Hence, if $\epsilon_1 \in (0,1)$ is chosen large enough 
		such that ${\rho}_{j} I - (1 - \epsilon_1)\lambda_{j}\tilde{P}_{j} \succeq 0$, then for all $\delta_j \in[0,\tilde{\delta}_j)$ with $\tilde{\delta}_j \le \sqrt{\Vert {\rho}_{j} I - (1 - \epsilon_1)\lambda_{j}\tilde{P}_{j}\Vert}$, the right-hand side of inequality \eqref{eq:mu:rho:2} is negative semi-definite.
		This ensures that constraint \eqref{eq:lqr:ball:1} is satisfied.
		If we can construct $\gamma$, $L$, and $Q$ satisfying constraints \eqref{eq:lqr:ball:2}--\eqref{eq:lqr:ball:5}, the feasibility of SDP \eqref{eq:lqr:ball} is proved.
		
		By letting $\tilde{Q}_j := (\lambda_j \tilde{P}_j)^{-1}$ and $\hat{\gamma}_{j} := \tilde{\gamma}_j + \|\tilde{P}_j^{-1}\|\epsilon_2/\lambda_j$, it is evident that constraints \eqref{eq:lqr:ball:2}--\eqref{eq:lqr:ball:5} are automatically satisfied.
		Therefore, $( \hat{\gamma}_j, \lambda_j\tilde{P}_j, \lambda_j\tilde{Y}_j, \lambda_j\tilde{L}_j, \tilde{Q}_j)$ is a candidate solution of SDP \eqref{eq:lqr:ball} and one has from $\mathcal{I}_t^\ast \subseteq \mathcal{B}^{\delta_j} $ that SDP \eqref{eq:lqr:on}  is feasible for set $\mathcal{I}_t^\ast$.
		
		\textbf{Case 2: $\lambda_j \ge 1$.}
		
		When $\lambda_j \ge 1$,  letting $\tilde{P}_{j}^{\lambda_j} := \lambda_j\tilde{P}_{j}$,  $\tilde{Y}_{j}^{\lambda_j} := \lambda_j \tilde{Y}_{j}$, $\tilde{L}_{j}^{\lambda_j} := \lambda_j \tilde{L}_{j}$ and $\tilde{\gamma}_{j}^{\lambda_j} := \lambda_j \tilde{\gamma}_{j}$, in the following, we first show that $(\tilde{P}_{j}^{\lambda_j},\tilde{Y}_{j}^{\lambda_j},\tilde{L}_{j}^{\lambda_j},\tilde{\gamma}_{j}^{\lambda_j})$ is a candidate solution of problem \eqref{eq:lqr:peter:ori} and then use these variables to construct a candidate solution for SDP \eqref{eq:lqr:ball}.

		Since $\lambda_j \ge 1$, constraints \eqref{eq:lqr:peter:ori:2}--\eqref{eq:lqr:peter:ori:4} are automatically satisfied.
		Multiplying both sides of \eqref{eq:py1} with $\lambda_j$, one gets that
		\begin{equation*}
			\lambda_j[A_j~B_{\rm tr}]
			\left[
			\begin{matrix}
				\tilde{P}_{j}\\
				\tilde{Y}_{j}
			\end{matrix}
			\right] \cdot\tilde{P}_{j}^{-1}[\star]' - \lambda_j\tilde{P}_{j} \preceq -\lambda_j I \preceq -I
		\end{equation*}
		which implies that \eqref{eq:lqr:peter:ori:1} is satisfied.
		Hence $(\tilde{P}_{j}^{\lambda_j}, \tilde{Y}_{j}^{\lambda_j},\tilde{L}_{j}^{\lambda_j},\\ \tilde{\gamma}_{j}^{\lambda_j})$ is a candidate solution to SDP \eqref{eq:lqr:peter:ori}.
		
		We now consider $(\tilde{P}_{j}^{\lambda_j},\tilde{Y}_{j}^{\lambda_j},\tilde{L}_{j}^{\lambda_j},\tilde{\gamma}_{j}^{\lambda_j})$ and show that there exists some $\mu_j^{\lambda_j} \in (0,1)$, such that constraint \eqref{eq:lqr:ball:1} is satisfied at $(\mu_j^{\lambda_j}\tilde{P}_j^{\lambda_j}, \mu_j^{\lambda_j}\tilde{Y}_j^{\lambda_j})$ under some conditions on $\epsilon_1$ and $\delta_j$.
		Noting that inequality \eqref{eq:rho} becomes	\begin{align}\label{eq:mu:rho:biglamabda}
				\left[
				\begin{matrix}
					-\tilde{P}_{j}^{\lambda_j} - Z_{j}'Z_{j} & 0& -Z_{j}'\\
					0 & -\tilde{P}_{j}^{\lambda_j} & [\tilde{P}_{j}^{\lambda_j}~ \tilde{Y}_{j}^{\lambda_j\prime}] \\
					-Z_{j}& [\tilde{P}_{j}^{\lambda_j}~ \tilde{Y}_{j}^{\lambda_j\prime}]' & -I\\
				\end{matrix}
				\right] \preceq - {\rho}_{j} I. 
		\end{align}
		Multiplying both sides of inequality  \eqref{eq:mu:rho:biglamabda} by $\mu_j^{\lambda_j}$ and substituting it into constraint \eqref{eq:lqr:ball:1} yields \eqref{eq:epsilon:delta}, which is presented at the top of the next page.
		Since $\mu_j^{\lambda_j} \in(0,1)$, based on Schur complement lemma, if 
		\setcounter{equation}{30} 
		\begin{align}\label{eq:mu1}
			&~~-\mu_j^{\lambda_j} \rho_{j} I \!+\! \mu_j^{\lambda_j} (1 \!-\! \epsilon_1) \tilde{P}_{j}^{\lambda_j} \!+\! \delta_j^2 I \nonumber\\
			&~~- (1\!-\!\mu_j^{\lambda_j})Z_{j}'Z_{j}   \!-\! \frac{(1-\mu_j^{\lambda_j})^2}{-\mu_j^{\lambda_j} \rho_{j} + \mu_j^{\lambda_j} - 1} Z_{j}'Z_{j} \nonumber\\
			&\prec -\mu_j^{\lambda_j} \rho_{j} I + \mu_j^{\lambda_j} (1 - \epsilon_1) \tilde{P}_{j}^{\lambda_j} + \delta_j^2 I \preceq 0
		\end{align}
		then constraint \eqref{eq:lqr:ball:1} is satisfied.
		Therefore, if $\epsilon_1 \in (0,1)$ is chosen rendering $\rho_{j}I - (1- \epsilon_1)\tilde{P}_{j}^{\lambda_j} \prec 0$, then for $\delta_j \in [0,\tilde{\delta}_j]$ with $\tilde{\delta}_j \le \sqrt{\mu_j\Vert\rho_{j}I - (1- \epsilon_1)\tilde{P}_{j}^{\lambda_j}\Vert}$, inequality \eqref{eq:mu1} is satisfied.
		Moreover, since $\tilde{\delta}_j \le \delta$, we can find a positive constant $\bar{\delta} > 0$ such that inequality \eqref{eq:mu1} is satisfied and consequently \eqref{eq:epsilon:delta} is non-positive for all $\delta \in [0,\bar{\delta})$. This inequality guarantees the satisfaction of constraint \eqref{eq:lqr:on:1}. 
		
		Moreover, letting $\hat{\gamma}_{j}^{\lambda_j} :=  \tilde{\gamma}_j^{\lambda_j} +\epsilon_2\|\tilde{Q}_j^{\lambda_j}\|$ and $\tilde{Q}_j^{\lambda_j} := (\tilde{P}_j^{\lambda_j}\mu_j^{\lambda_j})^{-1}$, similar to the proof for the case $\lambda_j\in (0,1)$,  constraints \eqref{eq:lqr:ball:2}--\eqref{eq:lqr:ball:5} are automatically satisfied.
		Hence, $( \hat{\gamma}_j^{\lambda_j}, \mu_j^{\lambda_j}\tilde{P}_j^{\lambda_j}, \mu_j^{\lambda_j}\tilde{Y}_j^{\lambda_j}, \mu_j^{\lambda_j}\tilde{L}_j^{\lambda_j}, \tilde{Q}_j^{\lambda_j})$ is a solution to SDP \eqref{eq:lqr:ball}, indicating that for all $\delta \in [0, \bar{\delta})$, SDP \eqref{eq:lqr:on} is feasible.
		
		So far, we have established the feasibility of SDP \eqref{eq:lqr:on}.
		Let  $(\gamma^*(t), P^*(t), Y^*(t), L^*(t), Q^*(t))$ be an optimal solution of \eqref{eq:lqr:on} at time $t$ and $K(t) = Y^*(t)(P^*(t))^{-1}$.
		In the sequel, we prove the statements s1) and s2). 
		
		\emph{ii) Proof of statements s1) and s2).}

		Since $Z_j = [A_j~ B_{\rm tr}]' \in \mathcal{I}_t^\ast,\forall t \in [t_s + 1,t_{s +1}]$,
		according to \cite[Theorem 1]{bisoffi2022data},  the feasibility of \eqref{eq:lqr:on} implies the following inequality
		\begin{equation}\label{eq:stable:K}
			[A_j + B_{\rm tr}K(t)]\cdot P^*(t)[\star]' - P^*(t)  \preceq ( \epsilon_1-1)P^*(t).
		\end{equation} 
		This further indicates that  $K(t)$ stabilizes the subsystem $(A_j,B_{\rm tr})$ for all $t \in [t_s + 1,t_{s +1}]$, which proves statement s1).
		
		To show statement s2), we recall that $\tilde{P}_{j}$ and $\tilde{K}_{j}$ correspond to the unique LQR solutions for the subsystem  $(A_{j},B_{\rm tr})$.
		Since $\gamma^*(t) \le \hat{\gamma}_{j}$ if $\lambda_j \in (0,1)$ with $\hat{\gamma}_{j} = \lambda_{j} \tilde{\gamma}_{j} + \epsilon_2 \|\tilde{P}_{j}^{-1}\|/\lambda_{j}$ (or $\gamma^*(t) \le \hat{\gamma}_{j}^{\lambda}$ if $\lambda_j\ge 1$ with $\hat{\gamma}_{j}^{\lambda_j}  = \mu_{t_i}^{\lambda_j} \tilde{\gamma}_{j}^{\lambda_j} + \epsilon_2 \|(\tilde{P}_{j}^{{\lambda_j}})^{-1}\|/\mu_{j}^{\lambda_j}$), it follows that $\Vert (P^{*}(t))^{-1} \Vert \le \hat{\gamma}_{j}/\epsilon_2$ if $\lambda_j \in(0,1)$ (or $\Vert (P^{*}(t))^{-1} \Vert \le \hat{\gamma}_{j}^{\lambda}/\epsilon_2$ if $\lambda_j \ge1$).
		Hence, $P^*(t) \succeq \epsilon_2/\hat{\gamma}_{j} I$ if $\lambda_j \in(0,1)$ (or $P^*(t) \succeq \epsilon_2/\hat{\gamma}_{j}^{\lambda} I$ if $\lambda_j \ge 1$).
		It can be deduced from \eqref{eq:lqr:on:3}--\eqref{eq:lqr:on:4} that 
		\begin{align*}
			\Vert K(t)\Vert &\le \sqrt{{\rm Tr}(K(t)I K'(t))} \\
			&\le \sqrt{{\rm Tr}(K(t)P^*(t)K'(t))\max\{\hat{\gamma}_{j},\hat{\gamma}_{j}^{\lambda_j}\}/\epsilon_2}\\
			&\le \sqrt{{\rm Tr}(L^*(t))\max\{\hat{\gamma}_{j},\hat{\gamma}_{j}^{\lambda_j}\}/\epsilon_2} \\
			&\le  \sqrt{\max\{\hat{\gamma}_{j},\hat{\gamma}_{j}^{\lambda_j}\}^2/\epsilon_2}.
		\end{align*} 
		Consequently, statement s2) can be proven by setting $\kappa := \max_{j \in \mathcal{M}} \sqrt{\max\{\hat{\gamma}_{j},\hat{\gamma}_{j}^{\lambda_j}\}^2/\epsilon_2}$.
	\end{pf}
	
	\begin{remark}[Complexity]\label{rmk:complexity}
		The proposed controller requires computing SDPs \eqref{eq:intersection} and \eqref{eq:lqr:on} at each time.
		According to \cite[Section G]{sznaier2020control}, the complexity of solving SDP \eqref{eq:intersection} is roughly $\mathcal{O}((n_x + n_u)^3(2n_x + n_u)^3)$ and that of solving SDP \eqref{eq:lqr:on} is roughly $\mathcal{O}(n_x^3(n_x + n_u)^3)$.
		One can further consolidate these two SDPs into a single SDP as follows
		\begin{subequations}
			\footnotesize
			\begin{align}
				\min_{\gamma,\beta,  \tau_1, \tau_2,\atop P, Y, L, Q} & ~~\gamma \label{eq:lqr:on:a}\\
				{\rm s.t. }  ~~~ & \left[
				\begin{matrix}
					P - \beta I\!\!&0 \!\!\!& 0\!\!\!& 0\\
					0 \!\!& -P \!\!\!& -Y' \!\!\!& 0\\
					0 \!\!& -Y \!\!\!& 0 \!\!\!& Y\\
					0 \!\!&0 \!\!\!& Y' \!\!\!& P
				\end{matrix}
				\right] 
				- \tau_1 
				\left[
				\begin{matrix}
					-\mathbf{C}_t \!&\! -\mathbf{B}_t' \!&\!\! 0\\
					-\mathbf{B}_t \!&\! -\mathbf{A}_t \!&\!\! 0\\
					0 \!&\! 0 \!&\!\! 0
				\end{matrix}
				\right] 
				\nonumber\\
				&- 
				\tau_2 
				\left[
				\begin{matrix}
					-\mathbf{C}^{\delta} & -Z_{\rm tr}' & 0\\
					0 & -I & 0\\
					0 & 0 & 0
				\end{matrix}
				\right] \succeq 0 \label{eq:lqr:on:b}\\
				& \beta >0,~~\tau_1 \ge 0,~~ \tau_2 \ge 0, ~~P \succ 0\\
				&\left[
				\begin{matrix}
					L & Y\\
					Y'& P
				\end{matrix}
				\right] \succeq 0\label{eq:lqr:on:d}\\
				&\left[
				\begin{matrix}
					Q & I\\
					I & P
				\end{matrix}
				\right] \succeq 0 \label{eq:lqr:on:e}\\
				&{\rm Tr}(P) + {\rm Tr}(L) + \epsilon \Vert Q\Vert\le \gamma. \label{eq:lqr:on:f}
			\end{align}
		\end{subequations}
		In this manner, the computational complexity reduces to $\mathcal{O}(n_x^3(n_x + n_u)^3)$, which is smaller than $\mathcal{O}(n_x^3(n_x(n_u + 1) + n_u)^3)$ in \cite{rotulo2021online}.
	\end{remark}
\begin{remark}
	[Noisy data]
	The proposed method can be directly extended to handle noisy data. Instead of considering the ideal system \eqref{eq:sys:ideal}, we consider a system with noise, i.e., $x(t + 1) = A x(t) + B u(t) + w(t)$, where the process noise $w(t)$ satisfies $\Vert w(t)\Vert \le \bar{w}$ for all $t \in \mathbb{N}$.
	In this setting, the value of $\delta$ is affected by both the attack matrix and the noise level $\bar{w}$.
	During the online step, the matrix $\mathbf{C}_t$ in \eqref{eq:ellipsoid:t:bc} can be replaced by $\mathbf{C}_t = x_p(t) x_p'(t) - \bar{w}^2I$ to compensate for noise.
\end{remark}
	Thus far, we have established the feasibility for SDP \eqref{eq:lqr:on} provided that certain conditions on the attacking power $\delta$ and the parameter $\epsilon_1$ are met.
	In addition, matrix $K(t)$ defined in \eqref{eq:Kk} stabilizes the subsystem  $(A_{\sigma(t_s)},B_{\rm tr})$ for $t \in [t_{s} + 1, t_{s+1}]$.
	However, observing at time $t=t_{s+1}$ that the dynamics switches to $(A_{\sigma(t_{s+1})}, B_{\rm tr})$ while the matrix $K(t)$ is designed to stabilize $(A_{\sigma(t_{s})},B_{\rm tr})$.
	Consequently, there is a mismatch between the controller gain $K(t)$ and the activated mode $\sigma(t_{s+1})$, which  may lead to divergence of the state.
	Hence, in the following,  conditions on the attacker's switching frequency is derived to ensure the stability of system \eqref{eq:sys}.
	\begin{remark}[Parameters $\epsilon_1$ and $\epsilon_2$]\label{rmk:epsilon}
		It has been shown in the proof of Theorem \ref{thm:lqr} that a larger $\epsilon_1$ enables the feasibility of SDP \eqref{eq:lqr:on} under a larger attack energy bound $\delta$.
		The largest bound is achieved when $\epsilon_1 = 1$.
		In this case, constraint \eqref{eq:lqr:on:1} should be modified into a strict form, i.e., replacing $\preceq$ with $\prec$.
		On the other hand, inequalities \eqref{eq:stable:K} and \eqref{eq:alphaj} in the proof of Theorem \ref{thm:stability} imply that a smaller $\epsilon_1$ and a  larger $P^*(t)$ provides a faster convergence rate, which leads to better system's performance as we  illustrate through numerical examples in the ensuing section.
		The parameter $\epsilon_1$ trades off between the feasibility of  SDP \eqref{eq:lqr:on} and the system's performance.
		Furthermore, as discussed before, system's performance benefits from a larger $P^*(t)$, according to \eqref{eq:lqr:on:4} and \eqref{eq:lqr:on:5}, which can be realized by choosing a larger $\epsilon_2$.
	\end{remark}
	
	\begin{theorem}\label{thm:stability}
		Under Assumptions \ref{as:2:tre}---\ref{as:5:ctrl}, let $U_0$, $X_0$, and $X_1$ be given in \eqref{eq:X0}.
		There exists constants $\bar{\delta} >0$, $\bar{\tau} \ge 2$ such that for all $\delta \in [0,\bar{\delta})$ and $\tau > \bar{\tau}$,
		the system \eqref{eq:sys} with Algorithm \ref{alg:ctrl} is exponentially stable, for arbitrary initial conditions $x_p(0)$ and $u_o(0)$.
	\end{theorem}
	\begin{pf}
		Consider the system \eqref{eq:sys} in any switching interval $[t_s,t_{s + 1}-1]$ with Algorithm \ref{alg:ctrl}. Let $j \in \mathcal{M}$ denote the subsystem selected by $\sigma(t_s)$, i.e., $j = \sigma(t)$ for all $t \in [t_s, t_{s + 1}-1]$.
		First, we demonstrate the convergence of the state for all $t \in [t_s + 2,t_{s + 1}]$ and any $s \in \mathbb{N}$. 
		By bounding the distance between $x(t_s + 1)$ and $x(t_s)$, we can achieve stability under mild conditions on the switching frequency and the attacking power.
		
		According to statement s1) of Theorem \ref{thm:lqr}, it follows that for all $t \in [t_s + 1, t_{s + 1}]$, SDP \eqref{eq:lqr:on} is feasible with an optimal solution $(\gamma^*(t), P^*(t), Y^*(t), L^*(t), Q^*(t))$, and $K(t) = Y^*(t)(P^*(t))^{-1}$ stabilizes $(A_{j}, B_{\rm tr})$. Therefore, there exist $P_j \succ 0$ and $\alpha_j \in (0,1)$ 
		 such that 
		\begin{equation}\label{eq:alphaj}
			(A_j + B_{\rm tr}K(t))'P_j(A_j + B_{\rm tr}K(t)) - P_j \preceq -\alpha_j I
		\end{equation}
		holds for all $t \in [t_s+1, t_{s + 1}]$.
		
		We construct a Lyapunov function $V_j(t) = x_p'(t)P_jx_p(t)$, which satisfies the inequalities
		{\setlength{\abovedisplayskip}{5pt}
		\setlength{\belowdisplayskip}{5pt}
		\begin{equation}\label{eq:xVx}
			\underline{\lambda}_{P} \Vert x_p(t) \Vert^2 \le V_{j}(t) \le \bar{\lambda}_{P} \Vert x_p(t) \Vert^2
		\end{equation}}
		where $\underline{\lambda}_{P} := \min_{j \in \mathcal{M}} \underline{\lambda}_{P_j}$ and $\bar{\lambda}_{P} := \max_{j \in \mathcal{M}} \bar{\lambda}_{P_j}$.

		For $t \in [t_s + 1,t_{s + 1}-1]$, it can be deduced  that $V_j(t + 1) \le (1 - \alpha_j/\bar{\lambda}_P) V_j(t)$.
		By utilizing \eqref{eq:xVx}, we let $c_j = \sqrt{\bar{\lambda}_P/\underline{\lambda}_P}$ and $\bar{\alpha}_j = \sqrt{1 - \alpha_j/\bar{\lambda}_P}$, arriving at $\Vert x_p(t)\Vert  \le c_j \bar{\alpha}_j^{t - t_s-1} \Vert x_p(t_s+1)\Vert$ for $t \in [t_s + 2,t_{s + 1}]$.
		
		Furthermore, observing from statement s2) in Theorem \ref{thm:lqr} that $\Vert x_p(t_{s}+1)\Vert \le (\Vert A_{j}\Vert + \kappa\Vert B_{\rm tr}\Vert)\Vert x_p(t_s)\Vert$ holds for $s\in\mathbb{N}_+$.
		Noting for $t = t_0 = 0$, since $u_o(0)$ is bounded, there exists a constant $C_0$ such that $\Vert x(1)\Vert \le C_0\Vert x(0)\Vert$.
		Therefore, by defining $C_2 := \max\{C_0,C_1\}$ with $C_1 :=\max_{j \in \mathcal{M}} {\Vert A_{j}\Vert + \kappa\Vert B_{\rm tr}\Vert}$, we have $\Vert x_p(t_{s}+1)\Vert \le C_2\Vert x_p(t_s)\Vert$ for $s \in \mathbb{N}$.
		
		Let $\hat{c} :=\max_{j \in \mathcal{M}} c_j$ and $\hat{\alpha} := \max_{j \in \mathcal{M}} \bar{\alpha}_j$. Combining above results yields for any $t \in [t_s + 1, t_{s + 1}]$ that $\Vert x_p(t)\Vert   \le C_2 c_j \bar{\alpha}_j^{t - t_s-1} \Vert x_p(t_s)\Vert
		\le {C_2\hat{c}}/{\hat{\alpha}}\hat{\alpha}^{t - t_s}\Vert x_p(t_s)\Vert$.
		Through telescoping on $t$ and recursively expanding on $s$, it follows for all $t\in\mathbb{N}$ that
		{\setlength{\abovedisplayskip}{5pt}
		\setlength{\belowdisplayskip}{5pt}
		\begin{align}
			\Vert x_p(t)\Vert    	&\le \Big(\frac{C_2\hat{c}}{\hat{\alpha}}\Big)^{s + 1}\hat{\alpha}^{t - t_0}\Vert x_p(t_0)\Vert\label{eq:adt}\\
			&\le
			\underbrace{\Big(\frac{C_2\hat{c}}{\hat{\alpha}}\Big)^{\upsilon}}_{:= C_3} \Big[\Big(\frac{C_2\hat{c}}{\hat{\alpha}}\Big)^{\!\frac{1}{\tau}}\hat{\alpha}\Big]^{t - t_0}\Vert x_p(t_0)\Vert \label{eq:c3}
		\end{align}}
		where the second inequality comes from Assumption \ref{as:dwell} with $s + 1 = N_{\sigma}(t_0,t) \le \upsilon + (t - t_0)/\tau$ for some $\upsilon \ge 0$.
		Hence, when $\tau > \bar{\tau}$ with $\bar{\tau} := (\ln\hat{\alpha} - \ln{C_2} -\ln{\hat{c}})/\ln\hat{\alpha}$
		then $0<(C_2\hat{c}/\hat{\alpha})^{{1}/{\tau}}\hat{\alpha} <1$.
		Consequently, upon defining $\beta :=(C_2\hat{c}/\hat{\alpha})^{{1}/{\tau}}\hat{\alpha}$, we arrive at $\Vert x_p(t)\Vert \le C_3 \beta^{t}\Vert x_p(0)\Vert$
		with the constant $C_3$ in \eqref{eq:c3}. This completes the proof.
	\end{pf}

				\section{Numerical Examples}
				In this section, we compare the proposed scheme with the system identification-based approach in \cite{wu2019optimalswitching} and the data-driven method in \cite{rotulo2021online} through two numerical examples.
				The following simulations were run on a Lenovo laptop with a 14-core 2.3GHz i7-12700H processor. 
				{\setlength{\textfloatsep}{5pt}
				\begin{figure}
					\centering
					\includegraphics[width=8cm]{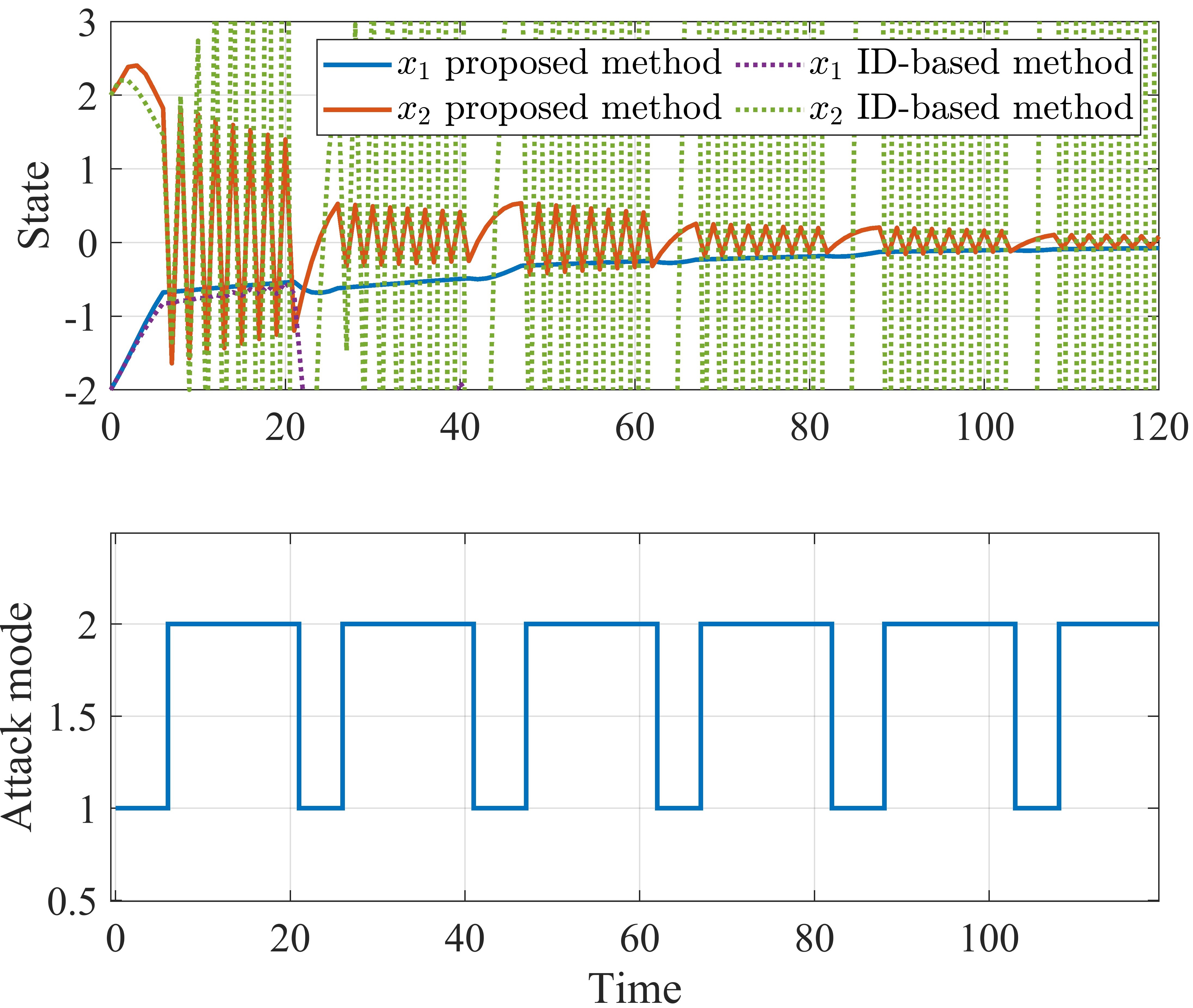}\\
					\caption{System performance for Section \ref{sec:num:1}. Top: System trajectories under the proposed resilient controller and the ID-based method in \cite{wu2019optimalswitching}; Bottom: Switching times of attacks.}\label{fig:com_wu_2_xc}
					\centering
				\end{figure}}
			{\setlength{\textfloatsep}{5pt}
				\begin{figure}
					\centering
					\includegraphics[width=8cm]{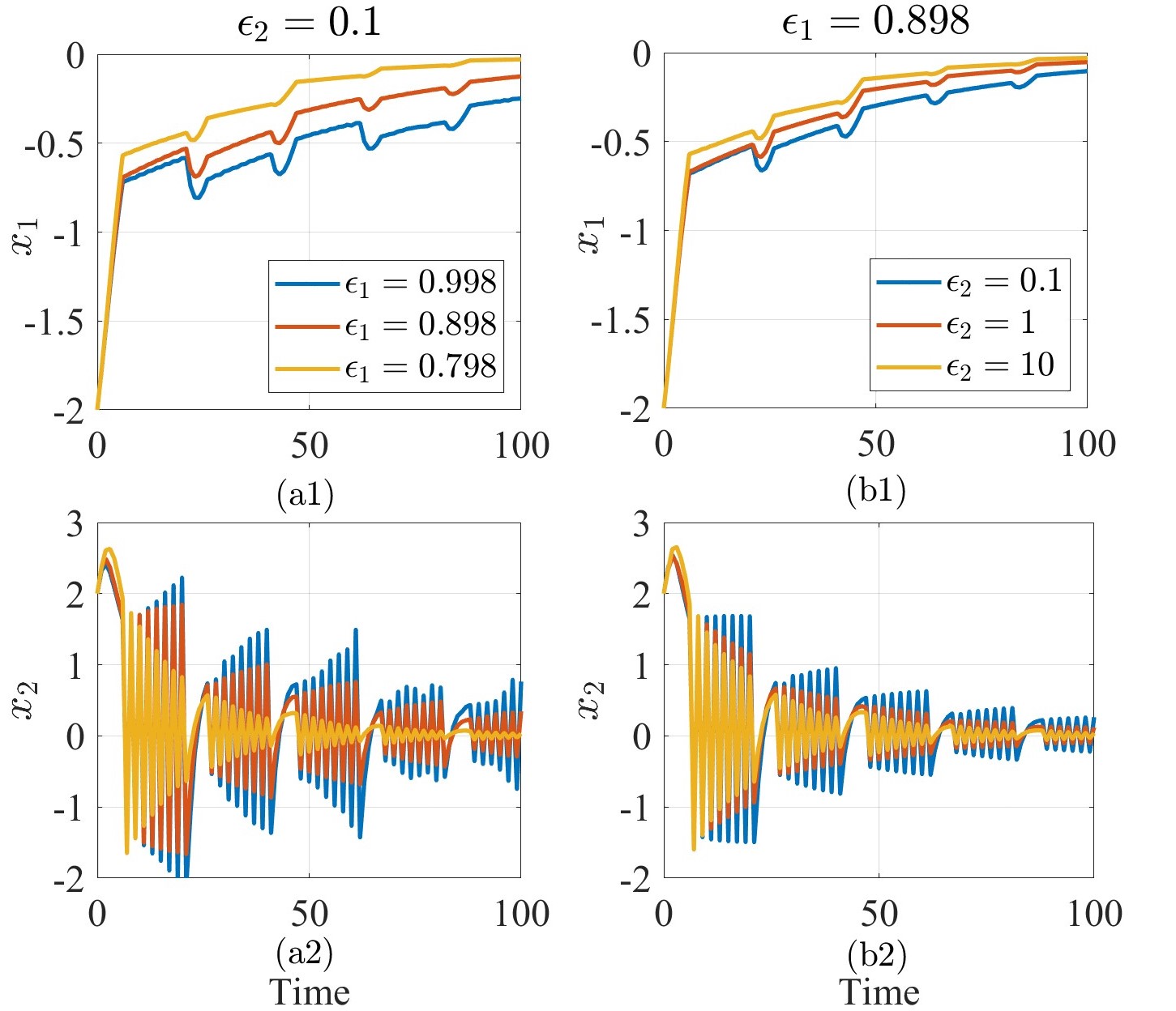}\\
					\caption{System's performances under different $\epsilon_1$ and $\epsilon_2$.}\label{fig:com_wu_2_epsilon}
					\centering
				\end{figure}}
				
				\subsection{Power Generator}\label{sec:num:1}
				Taking the sampling time to be $T_s = 0.1$s, we consider a discretized version of the linearized model of the normalized swing equation in \cite{wu2019optimalswitching}, where
				{\setlength{\abovedisplayskip}{5pt}
				\setlength{\belowdisplayskip}{5pt}
				\begin{align}\label{eq:num:1:sys}
					A_{\rm tr} = \left[
					\begin{matrix}
						0& 1\\
						-1& -1
					\end{matrix}
					\right],~B_{\rm tr} = \left[
					\begin{matrix}
						0& 0\\
						1& 1
					\end{matrix}
					\right],~ K = \left[
					\begin{matrix}
						-1& 0\\
						0& -2
					\end{matrix}
					\right].
				\end{align}}
				
				For simplicity, we assume that the attacker knows the matrices $A_{\rm tr}, B_{\rm tr} $, $ K$, and generates the attacking matrices using the method in \cite{wu2019optimalswitching}.
				For a fair comparison, we multiply the obtained attack matrices by $1.2$ (i.e., increasing the power of attacks), yielding attacking matrices
				$B_{\rm tr}D_a^1K_a^1 = \left[
				\begin{matrix}
					-0.0232 &  -0.0103\\
					-0.4567  & -0.2025
				\end{matrix}
				\right]$ and $ 
				B_{\rm tr}D_a^2K_a^2 =\left[
				\begin{matrix}
					-0.0107 &   -0.0929\\
					-0.2110 &  -1.8266
				\end{matrix}
				\right]$.
				It is easy to verify that $\Vert B_{\rm tr}D_a^j K_a^j \Vert \le 1.841$ for $j=1,2$. 
				The switching signal $\sigma$ is given in the bottom panel of Fig. \ref{fig:com_wu_2_xc}.
				The resilient controller designed based on a least-squares system identification (ID) method in \cite{wu2019optimalswitching} is given by $u_o(t) = K_{ID} x_p(t)$ with
				 $K_{ID} = \left[
				\begin{matrix}
					1.6516  &  0.2865\\
					-0.0997  &  1.9777
				\end{matrix}
				\right].$
				
				Setting $T = 15$, we collect an offline input-state trajectory by applying a sequence of inputs $u(t)$ uniformly generated from $[-0.3, 0.3]$ to the healthy system $(A_{\rm tr}, B_{\rm tr})$. 
				Taking $\delta = 0.125$, $\epsilon_1 =0.898$, and $\epsilon_2 = 1$, over a simulation horizon of $120$ time steps, the top panel of Fig. \ref{fig:com_wu_2_xc} compares the system's performance under the proposed method in Alg. \ref{alg:ctrl} (solid line) and the ID-based resilient controller in \cite{wu2019optimalswitching} (+ marked dashed line), showcasing the effectiveness of the proposed method.
				Moreover, the system's performance under different $\epsilon_1$ and $\epsilon_2$ values in SDP \eqref{eq:lqr:on} is depicted in Fig. \ref{fig:com_wu_2_epsilon}.
				It is shown that smaller $\epsilon_1$ and larger  $\epsilon_2$ lead to better system's performance.
				However, when $\epsilon_1 \le 0.75$, SDP \eqref{eq:lqr:on} becomes infeasible for some $t$.

				\subsection{Aircraft Engine System}\label{sec:num:2}
				In the second example, we consider our approach to stabilizing fault tolerant systems. 
				Specifically, we apply the proposed controller to an F-$404$ aircraft engine system subject to system fault, see e.g., \cite{rotulo2021online}.
				We consider a discretized linearized version of the system with a sampling period of $0.1$ s, and the system matrices are given by 	
				{\setlength{\abovedisplayskip}{5pt}
				\setlength{\belowdisplayskip}{5pt}
				\begin{align*}
					A_{\rm tr} = \left[
					\begin{matrix}
						0.867 & 0& 0.202\\
						0.015 &0.961 &-0.032\\
						0.026& 0& 0.803\\
					\end{matrix}
					\right],~B_{\rm tr} = \left[
					\begin{matrix}
						0.011& 0\\
						0.014 &-0.039\\
						0.009& 0\\
					\end{matrix}
					\right].
				\end{align*} }

\begin{figure}
					\centering
					\includegraphics[width=8cm]{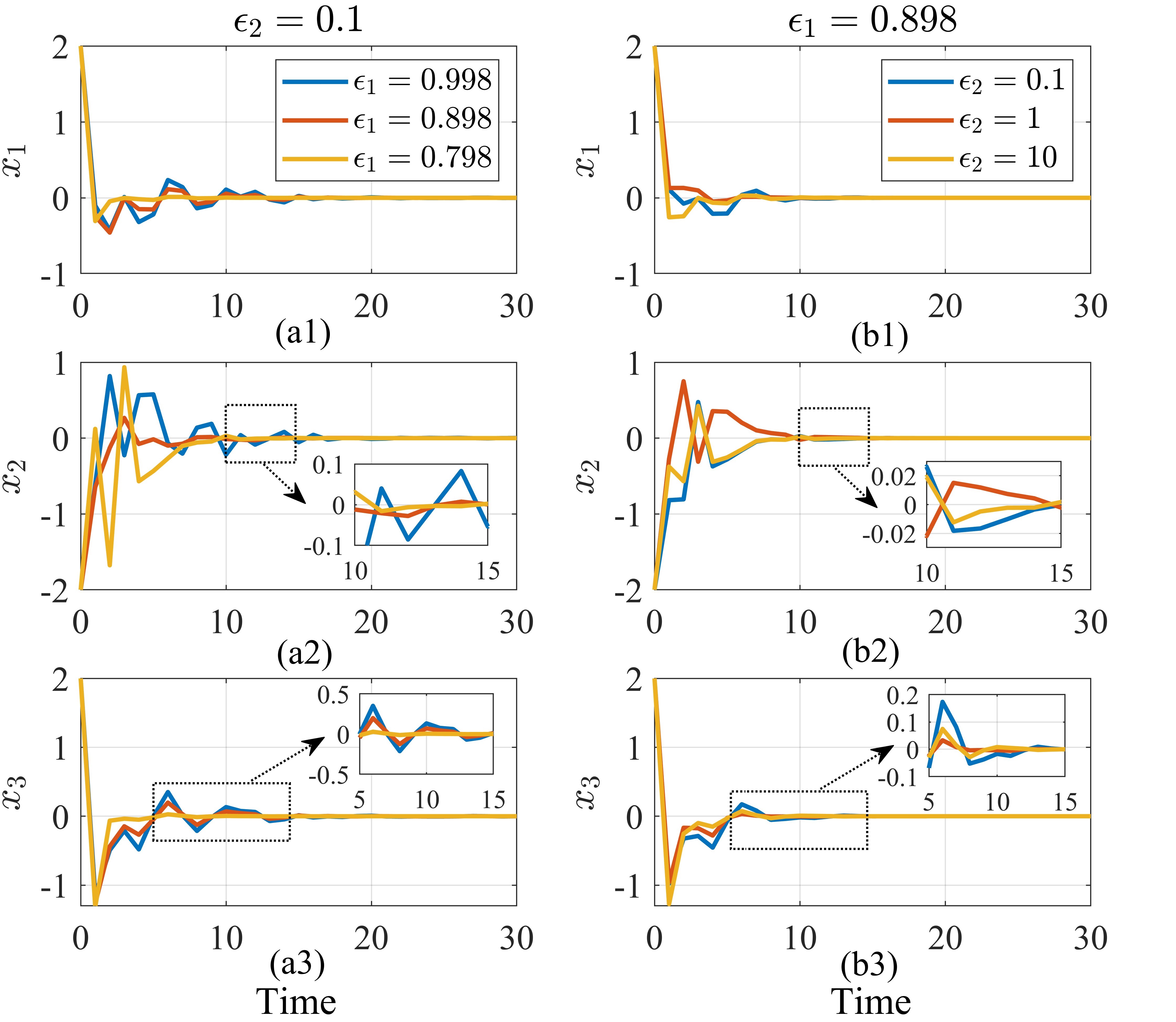}\\
					\caption{System's performances under different $\epsilon_1$ and $\epsilon_2$.}\label{fig:com_wmo_3_epsilon}
					\centering
				\end{figure}

				Similarly to the previous section, we collect an input-state trajectory of length $T=21$ by stimulating the system such that ${\rm rank}(W_0) = n_x + n_u$ holds. 
				We then run the system online, where unknown faults are characterized by changes in the system matrix $A_{\rm tr}$, resulting in $\tilde{A} = A_{\rm tr} + \omega(\sigma(t))D$ with
				{\setlength{\abovedisplayskip}{5pt}
				\setlength{\belowdisplayskip}{5pt}
				\begin{align*}
					D = \left[
					\begin{matrix}
						0.075 & 0 & 0\\
						0.5 & 1 & 0\\
						0 &0  &-0.75
					\end{matrix}
					\right],~\omega(\sigma(t)) = \begin{cases}
						0.7,& \sigma(t) = 1\\
						0.6, & \sigma(t) = 2\\
						-0.5, & \sigma(t) = 3
					\end{cases}
				\end{align*}  }
				where $\Vert \omega(\sigma(t))D\Vert \le 0.783$.
				The switching signal $\sigma(t)$ is depicted in Fig \ref{fig:com_wu_3_xc}(d).
				Letting $\delta = 0.14$, $\epsilon_1 = 0.998$, and $\epsilon_2 = 1$, under the fault switching signal presented in Fig. \ref{fig:com_wu_3_xc}(d), system trajectories under the proposed Alg. \ref{alg:ctrl} (+ marked dotted line), the data-driven method in \cite{rotulo2021online} (solid line), and a time-invariant feedback gain (dashed line) are depicted in Fig, \ref{fig:com_wu_3_xc}(a)--(c), which further corroborates the effectiveness of the proposed method.
				The system's performance under different $\epsilon_1$ and $\epsilon_2$ values is depicted in Fig. \ref{fig:com_wmo_3_epsilon}.
				The average computation time per iteration for the ID-based method in \cite{wu2019optimalswitching}, the method in \cite{rotulo2021online}, and the proposed method are $10^{-4}$s, $0.3394$s and $0.6133$s, respectively.
				Furthermore, by consolidating the SDPs \eqref{eq:intersection} and \eqref{eq:lqr:on} into single SDP, the computational time decreases to $0.2201$s, verifying Remark \ref{rmk:complexity}.
				{\setlength{\textfloatsep}{2pt}
				\begin{figure}
					\centering
					\includegraphics[width=8cm]{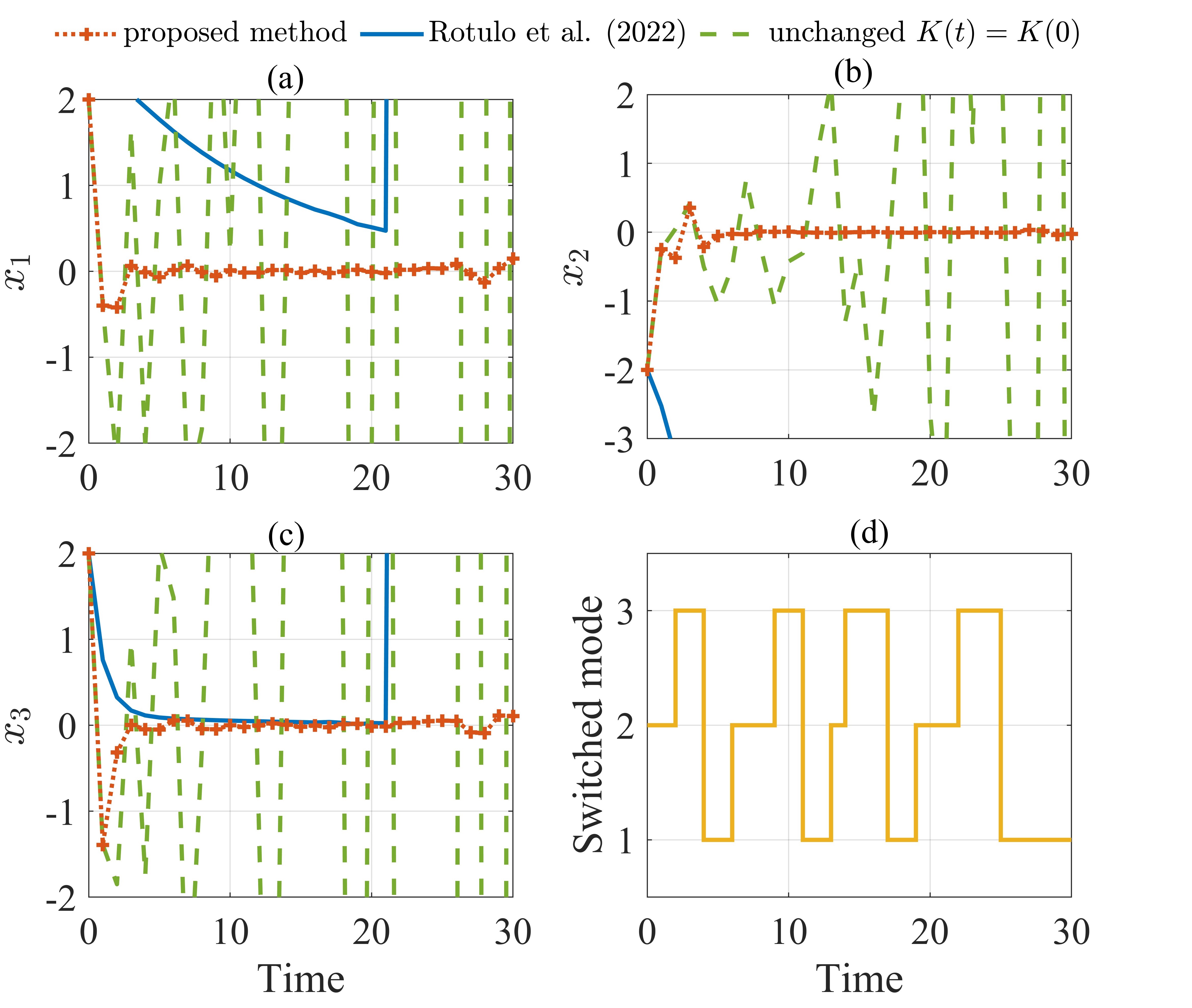}\\
					\caption{System's performance for Section \ref{sec:num:2}. Panels (a)--(c): state trajectories under the proposed Alg. \ref{alg:ctrl} (+ marked dotted line), data-driven method in \cite{rotulo2021online} (solid line), and time-invariant controller (dashed line); Panel (d): Switching times.}\label{fig:com_wu_3_xc}
					\centering
				\end{figure}}
				
				\section{Conclusions}
				This paper has investigated the stabilization problem of unknown linear systems under FDI attacks on the actuator channels using input-state data. A general FDI attack model was proposed by constraining the switching frequency and the attack power. Based on the model, the FDI-corrupted system can be modeled as a linear switched system.
				A method for minimally representing the system was developed, by combining online observations from the attacked system with offline data from the healthy system to compute a minimum-volume matrix ellipsoid by solving an SDP. In reminiscent of data-driven LQR, a data-based SDP was formulated and solved, from which a data-driven controller was devised. 
				Both SDPs' feasibility and exponential stability were established
				 Numerical examples were provided to illustrate the efficacy of the proposed controller.
				\bibliographystyle{plainnat}
				\bibliography{ddFDI}

\begin{thebibliography}{33}
\providecommand{\natexlab}[1]{#1}
\providecommand{\url}[1]{\texttt{#1}}
\expandafter\ifx\csname urlstyle\endcsname\relax
  \providecommand{\doi}[1]{doi: #1}\else
  \providecommand{\doi}{doi: \begingroup \urlstyle{rm}\Url}\fi

\bibitem[Anand and Teixeira(2022)]{anand2022risk}
S.~C. Anand and A.~M.~H. Teixeira.
\newblock Risk-averse controller design against data injection attacks on
  actuators for uncertain control systems.
\newblock In \emph{Proc. Amer. Control Conf.}, pages 5037--5042, Atlanta, GA,
  USA, June, 8-10, 2022.

\bibitem[Bai et~al.(2017)Bai, Pasqualetti, and Gupta]{bai2017data}
C.~Bai, F.~Pasqualetti, and V.~Gupta.
\newblock Data-injection attacks in stochastic control systems: Detectability
  and performance tradeoffs.
\newblock \emph{Automatica}, 82:\penalty0 251--260, Aug. 2017.

\bibitem[Bisoffi et~al.(2022)Bisoffi, De~Persis, and Tesi]{bisoffi2022data}
A.~Bisoffi, C.~De~Persis, and P.~Tesi.
\newblock Data-driven control via {P}etersen’s lemma.
\newblock \emph{Automatica}, 145:\penalty0 110537, Nov. 2022.

\bibitem[C{\'{a}}rdenas et~al.(2008)C{\'{a}}rdenas, Amin, and
  Sastry]{SminSecure}
A.~C{\'{a}}rdenas, S.~Amin, and S.~Sastry.
\newblock Secure control: {T}owards survivable cyber-physical systems.
\newblock In \emph{Proc. Int. Conf. Distrib. Comput. Syst. Wksp.}, pages
  495--500, Beijing, China, June, 16-20, 2008.

\bibitem[Cheng et~al.(2019)Cheng, Yang, Chen, Qi, and Shi]{cheng2019event}
P.~Cheng, Z.~Yang, J.~Chen, Y.~Qi, and L.~Shi.
\newblock An event-based stealthy attack on remote state estimation.
\newblock \emph{IEEE Trans. Autom. Control}, 65\penalty0 (10):\penalty0
  4348--4355, Oct. 2019.

\bibitem[{Coulson} et~al.(2019){Coulson}, {Lygeros}, and
  {Dörfler}]{Coulson2019data}
J.~{Coulson}, J.~{Lygeros}, and F.~{Dörfler}.
\newblock Data-enabled predictive control: {I}n the shallows of the {DeePC}.
\newblock In \emph{Proc. Eur. Control Conf.}, pages 307--312, Naples, Italy,
  June, 25-28, 2019.

\bibitem[{De Persis} and Tesi(2015)]{PersisInput}
C.~{De Persis} and P.~Tesi.
\newblock Input-to-state stabilizing control under denial-of-service.
\newblock \emph{IEEE Trans. Autom. Control}, 60\penalty0 (11):\penalty0
  2930--2944, Nov. 2015.

\bibitem[{De Persis} and {Tesi}(2020)]{persis2020data}
C.~{De Persis} and P.~{Tesi}.
\newblock Formulas for data-driven control: {S}tabilization, optimality, and
  robustness.
\newblock \emph{IEEE Trans. Autom. Control}, 65\penalty0 (3):\penalty0
  909--924, Mar. 2020.

\bibitem[{De Persis} and Tesi(2021)]{depersis2021lowcomplexity}
C.~{De Persis} and P.~Tesi.
\newblock Low-complexity learning of linear quadratic regulators from noisy
  data.
\newblock \emph{Automatica}, 128:\penalty0 109548, June, 2021.

\bibitem[Fawzi et~al.(2014)Fawzi, Tabuada, and Diggavi]{fawzi2014secure}
H.~Fawzi, P.~Tabuada, and S.~Diggavi.
\newblock Secure estimation and control for cyber-physical systems under
  adversarial attacks.
\newblock \emph{IEEE Trans. Autom. Control}, 59\penalty0 (6):\penalty0
  1454--1467, Jan. 2014.

\bibitem[Feng and Tesi(2017)]{FengResilient}
S.~Feng and P.~Tesi.
\newblock Resilient control under denial-of-service: {R}obust design.
\newblock \emph{Automatica}, 79:\penalty0 42--51, 2017.

\bibitem[Hashemi and Ruths(2022)]{hashemi2022co}
N.~Hashemi and J.~Ruths.
\newblock Co-design for resilience and performance.
\newblock \emph{IEEE Trans. Control Netw. Syst.}, Dec. 2022.
\newblock \doi{10.1109/TCNS.2022.3229774}.

\bibitem[Hou et~al.(2022)Hou, Sun, Yang, and Pang]{hou2022deep}
F.~Hou, J.~Sun, Q.~Yang, and Z.~Pang.
\newblock Deep reinforcement learning for optimal denial-of-service attacks
  scheduling.
\newblock \emph{Sci. China Inf. Sci.}, 65\penalty0 (6):\penalty0 162201, June,
  2022.

\bibitem[Kang and You(2023)]{kang2023minimum}
Shubo Kang and Keyou You.
\newblock Minimum input design for direct data-driven property identification
  of unknown linear systems.
\newblock \emph{Automatica}, 156:\penalty0 111130, 2023.

\bibitem[Kotidis and Schreft(2022)]{kotidis2022cyberattacks}
A.~Kotidis and S.~Schreft.
\newblock Cyberattacks and financial stability: {E}vidence from a natural
  experiment, 2022.
\newblock URL \url{http://dx.doi.org/10.17016/FEDS.2022.025}.

\bibitem[Krishnan and Pasqualetti(2021)]{krishnan2021data}
V.~Krishnan and F.~Pasqualetti.
\newblock Data-driven attack detection for linear systems.
\newblock \emph{IEEE Control Syst. Lett.}, 5\penalty0 (2):\penalty0 671--676,
  June, 2021.

\bibitem[Lee(2008)]{lee2008cyber}
E.~A. Lee.
\newblock Cyber physical systems: {D}esign challenges.
\newblock In \emph{Proc. IEEE Int. Symp. Object Compon.-Oriented Real-Time
  Distrib. Comput.}, pages 363--369, Orlando, FL, USA, May, 5-7, 2008.

\bibitem[Li et~al.(2023)Li, Wang, Sun, Wang, and Chen]{li2023data}
Y.~Li, X.~Wang, J.~Sun, G.~Wang, and J.~Chen.
\newblock Data-driven consensus control of fully distributed event-triggered
  multi-agent systems.
\newblock \emph{Sci. China Inf. Sci.}, 66\penalty0 (5):\penalty0 152202, May,
  2023.

\bibitem[Liu et~al.(2023{\natexlab{a}})Liu, Sun, Wang, Bullo, and
  Chen]{Liu2021data}
W.~Liu, J.~Sun, G.~Wang, F.~Bullo, and J.~Chen.
\newblock Data-driven resilient predictive control under denial-of-service.
\newblock \emph{IEEE Trans. Autom. Control}, 68\penalty0 (8):\penalty0
  4722--4737, Aug. 2023{\natexlab{a}}.

\bibitem[Liu et~al.(2023{\natexlab{b}})Liu, Sun, Wang, Bullo, and
  Chen]{liu2023learning}
W.~Liu, J.~Sun, G.~Wang, F.~Bullo, and J.~Chen.
\newblock Learning robust data-based {LQG} controllers from noisy data.
\newblock \emph{IEEE Trans. Autom. Control}, pages 1--13, May,
  2023{\natexlab{b}}.
\newblock \doi{10.1109/TAC.2024.3409749}.

\bibitem[Luo et~al.(2023)Luo, Zhao, and He]{luo2023secure}
X.~Luo, C.~Zhao, and J.~He.
\newblock Secure multi-dimensional consensus algorithm against malicious
  attacks.
\newblock \emph{Automatica}, 157:\penalty0 111224, Nov. 2023.

\bibitem[{McMillan}(2021)]{ransom2021robert}
R.~{McMillan}.
\newblock Ransomware attack affecting likely thousands of targets drags on.
\newblock
  \url{https://www.wsj.com/articles/ransomware-group-behind-meat-supply-attack-threatens-hundreds-of-new-targets-11625285071},
  2021.

\bibitem[Mo et~al.(2013)Mo, Hespanha, and Sinopoli]{mo2013resilient}
Y.~Mo, J.~P. Hespanha, and B.~Sinopoli.
\newblock Resilient detection in the presence of integrity attacks.
\newblock \emph{IEEE Trans. Signal Process.}, 62\penalty0 (1):\penalty0 31--43,
  Oct. 2013.

\bibitem[Murguia et~al.(2020)Murguia, Shames, Ruths, and
  Ne{\v{s}}i{\'c}]{murguia2020security}
C.~Murguia, I.~Shames, J.~Ruths, and D.~Ne{\v{s}}i{\'c}.
\newblock Security metrics and synthesis of secure control systems.
\newblock \emph{Automatica}, 115:\penalty0 108757, May, 2020.

\bibitem[Pasqualetti et~al.(2013)Pasqualetti, D{\"o}rfler, and
  Bullo]{pasqualetti2013attack}
F.~Pasqualetti, F.~D{\"o}rfler, and F.~Bullo.
\newblock Attack detection and identification in cyber-physical systems.
\newblock \emph{IEEE Trans. Autom. Control}, 58\penalty0 (11):\penalty0
  2715--2729, June, 2013.

\bibitem[Rotulo et~al.(2022)Rotulo, {De Persis}, and Tesi]{rotulo2021online}
M.~Rotulo, C.~{De Persis}, and P.~Tesi.
\newblock Online learning of data-driven controllers for unknown switched
  linear systems.
\newblock \emph{Automatica}, 145:\penalty0 110519, Nov. 2022.

\bibitem[Shi et~al.(2022)Shi, Feng, and Ishii]{shi2022quantized}
M.~Shi, S.~Feng, and H.~Ishii.
\newblock Quantized state feedback stabilization of nonlinear systems under
  denial-of-service.
\newblock \emph{Automatica}, 139:\penalty0 110180, May, 2022.

\bibitem[Sznaier(2020)]{sznaier2020control}
Mario Sznaier.
\newblock Control oriented learning in the era of big data.
\newblock \emph{IEEE Control Syst. Lett.}, 5\penalty0 (6):\penalty0 1855--1867,
  2020.

\bibitem[van Waarde and Camlibel(2021)]{van2021finsler}
H.~J. van Waarde and M.~K. Camlibel.
\newblock A matrix {F}insler’s lemma with applications to data-driven
  control.
\newblock In \emph{Proc. IEEE Conf. Decis. Control}, pages 5777--5782, Austin,
  TX, USA, Dec. 14-17, 2021.

\bibitem[Wakaiki et~al.(2019)Wakaiki, Cetinkaya, and
  Ishii]{wakaiki2019stabilization}
M.~Wakaiki, A.~Cetinkaya, and H.~Ishii.
\newblock Stabilization of networked control systems under {DoS} attacks and
  output quantization.
\newblock \emph{IEEE Trans. Autom. Control}, 65\penalty0 (8):\penalty0
  3560--3575, Oct. 2019.

\bibitem[{Willems} et~al.(2005){Willems}, {Markovsky}, {Rapisarda}, and {De
  Moor}]{willems2005note}
J.~C. {Willems}, I.~{Markovsky}, P.~{Rapisarda}, and B.~L.~M. {De Moor}.
\newblock A note on persistency of excitation.
\newblock \emph{Syst. Control Lett.}, 56\penalty0 (4):\penalty0 325--329, May,
  2005.

\bibitem[Wu et~al.(2020)Wu, Wang, Sun, and Chen]{wu2020optimal}
G.~Wu, G.~Wang, J.~Sun, and J.~Chen.
\newblock Optimal partial feedback attacks in cyber-physical power systems.
\newblock \emph{IEEE Trans. Autom. Control}, 65\penalty0 (9):\penalty0
  3919--3926, Mar. 2020.

\bibitem[Wu et~al.(2021)Wu, Sun, and Xiong]{wu2019optimalswitching}
G.~Wu, J.~Sun, and L.~Xiong.
\newblock Optimal switching attacks and countermeasures in cyber-physical
  systems.
\newblock \emph{IEEE Trans. Syst. Man Cybern. Syst.}, 51\penalty0 (8):\penalty0
  4825--4835, Aug. 2021.

\end{thebibliography}

			\end{document}